\documentclass[preprint,times]{elsarticle}
\def\acknow{%
\section*{Acknowledgements}

}

\usepackage{amsmath}
\usepackage{array,longtable}
\usepackage{textcomp}
\usepackage{bm} %
\usepackage{amsthm,amstext,amssymb,mathtools}
\usepackage[]{hyperref}
\usepackage[nameinlink]{cleveref}
\hypersetup{
    colorlinks,
    linkcolor={red!50!black},
    citecolor={blue!50!black},
    urlcolor={blue!80!black}
}
\usepackage[utf8]{inputenc}
\usepackage[T1]{fontenc}
\usepackage{appendix}
\usepackage{graphicx,sidecap}%
\usepackage[rawfloats]{floatrow}
\usepackage{grffile}
\usepackage{multirow,booktabs} %
\usepackage{verbatim}
\usepackage{todonotes}

\usepackage{fancyhdr}
\usepackage{empheq}
\usepackage{booktabs,colortbl}
\usepackage{siunitx}
\usepackage{xfrac}
\usepackage{capt-of}

\usepackage{graphicx}     %
\usepackage{subcaption}   %

\usepackage{tikz,pgfplots,pgfplotstable}
\pgfplotsset{compat=1.15}
\usetikzlibrary{matrix,arrows.meta,calc,shapes}
\tikzset{
  centered/.style = { align=center, anchor=center },
     empty/.style = { font=\sffamily\Large, centered, text width=2cm },
       box/.style = { font=\sffamily, fill=green, centered },
    result/.style = { font=\sffamily\scriptsize, fill=black!20, centered},
     arrow/.style = { very thick, color=red, ->, >=Triangle},
}

\usepackage{listings}

\newcommand{\vt}{{\bf v}}

\newcommand{\ut}{{\hat{\bf u}}}
\newcommand{\Jh}{\hat{J}}
\newcommand{\phit}{{\hat{\varphi}}}

\renewcommand{\div}{{\rm div}\,}

\newcommand{\tr}{{\rm tr}\,}

\newcommand{\D}{\mathbb{D}}

\newcommand{\I}{\mathbb{I}}

\newcommand{\Fb}{\hat{\mathbb{F}}}

\newcommand{\Bp}{\mathbb{B}_{\kappa_{p(t)}}}

\newcommand{\Fp}{\mathbb{F}_{\kappa_{p(t)}}}
\newcommand{\oldroyd}[1]{\stackrel{\triangledown}{#1}}

\renewcommand\Re{Re}
\newcommand\El{El}
\newcommand\St{S\!t}
\newcommand\We{W\!e}
\newcommand\Ca{C\!a}

\newtheorem*{remark}{Remark}

\begin{document}

\begin{frontmatter}
\title{A thermodynamically consistent Johnson--Segalman--Giesekus model: numerical simulation of the rod climbing effect}
\author[1]{Jakub~Cach} \ead{cach@mff.cuni.cz}
\author[2,1]{Patrick~E.~Farrell} \ead{patrick.farrell@maths.ox.ac.uk}
\author[1]{Josef~Málek\corref{c1}} \ead{josef.malek@mff.cuni.cz}
\author[1]{Karel~Tůma} \ead{ktuma@karlin.mff.cuni.cz}

\address[1]{Charles University, Faculty of Mathematics and Physics, Mathematical Institute, Sokolovsk\'{a} 83, 186 75 Prague, Czech Republic} 
\address[2]{University of Oxford, Mathematical Institute, Andrew Wiles Building, Woodstock Road, Oxford OX2 6GG, United Kingdom} 

\cortext[c1]{Corresponding author. The order of the authors is alphabetical.}

\begin{abstract}
{Viscoelastic rate-type fluids represent a popular class of non-Newtonian fluid models due to their ability to describe phenomena such as stress relaxation, non-linear creep, and normal stress differences. The presence of normal stress differences in a simple shear flow gives rise to forces acting in directions orthogonal to the primary flow direction. The rod climbing effect, i.e.~the rise of a fluid along a rod rotating about its axis, is associated with this phenomenon. Within the class of viscoelastic rate-type fluids that includes the Oldroyd-B and Giesekus models with Gordon--Schowalter convected derivatives, we show---by means of thermodynamical analysis and numerical simulations---that a thermodynamically consistent variant of the Johnson--Segalman model captures experimental data exceedingly well and is therefore superior to other models in this class, including the standard Johnson–Segalman model, which is widely used in engineering applications but is shown here to be incompatible with the second law of thermodynamics. We release a robust and computationally efficient higher-order finite-element implementation as open-source software on GitHub. The implementation is based on an arbitrary Lagrangian–Eulerian (ALE) formulation of the governing equations and is developed using the Firedrake library.}
\end{abstract}

\end{frontmatter}

\centerline{\emph{{\bf Dedicated to memory of Professor K.~R.~Rajagopal (November 24, 1950 -- March 20, 2025)}}
}

\section*{Keywords}
Viscoelasticity; Johnson--Segalman fluid; Oldroyd-B fluid; Giesekus fluid; %
Thermodynamic consistency; Rod climbing; Numerical simulation %

\section*{Highlights}
\begin{itemize}
\item The engineering Johnson--Segalman--Giesekus (JSG) model lacks a thermodynamic
consistency
\item A thermodynamically consistent variant of the JSG model is presented
\item High-order finite element implementation of ALE method demonstrated on rod climbing
\item Results for thermodynamically consistent JSG model agree better with experimental data
\end{itemize}

\section{Introduction}

This study aims to address the question of how well the class of standard first-order viscoelastic rate-type fluid models, including those of Oldroyd and Giesekus type, can describe the rod climbing phenomenon. To achieve this objective, we combine a transparent yet simple thermodynamic foundation for the fluids under consideration with advanced numerical simulations.

Rod climbing, i.e.~the rise of a fluid along a rod rotating about its axis, is a striking phenomenon that cannot be explained within the framework of Newtonian (Navier--Stokes) fluid mechanics. The rod climbing effect, also known as the Weissenberg effect, is associated with the presence of nonzero differences between the normal components of the Cauchy stress tensor in simple shear flows. This non-Newtonian phenomenon, commonly referred to as \emph{normal stress differences}, cannot be described by Newtonian fluids, nor by incompressible fluids with shear-rate or shear-stress dependent viscosity. In contrast, viscoelastic rate-type fluid models naturally account for normal stress differences and are therefore suitable candidates for reproducing the rod climbing phenomenon.

The first complete three-dimensional derivation of frame-indifferent viscoelastic rate-type fluid models was given by Oldroyd \cite{Oldroyd1950}, who introduced two distinct constitutive equations, referred to as Model~A and Model~B. Oldroyd already observed that these two models lead to qualitatively different predictions for free-surface flows. In particular, he noted that Model~A predicts a rise of the free surface near the outer stationary cylinder and a fall near the inner rotating cylinder, whereas Model~B predicts the opposite behavior, with the free surface tending to rise near the inner cylinder, i.e.\ exhibiting rod climbing. On this basis, Oldroyd explicitly identified Model~B as the variant capable of explaining the experimentally observed climbing effect \cite{Weissenberg1947,Weissenberg1948}. This predictive capability has played a central role in the widespread adoption of the latter model, called the Oldroyd-B fluid model, in both theoretical and experimental studies of viscoelastic fluids. The Oldroyd-B model employs the upper-convected time derivative to ensure frame indifference. However, other objective derivatives, such as the Jaumann--Zaremba or Gordon--Schowalter derivatives, are however also admissible from a purely kinematic standpoint.

While Oldroyd’s original derivation was carried out within a purely mechanical framework, it is natural to ask whether the Oldroyd-B model and other subsequently formulated three-dimensional viscoelastic models can be derived within a sound thermodynamic framework that guarantees compatibility with the laws of continuum thermodynamics. Several approaches have addressed this question; see, for example, Wapperom and Hulsen \cite{WapperomHulsen1998} and Grmela and \"{O}ttinger \cite{GrmelaOttingerGENERIC}. 

Here, we follow the thermodynamic approach proposed by Rajagopal and Srinivasa \cite{RajagopalSrinivasa2000,RajagopalSrinivasa2004}, which is built on a combination of two key concepts that provide the derived models with a clear physical interpretation. The first concept is associated with the notion of evolving natural configurations, which decompose the total deformation of the body into an elastic part and a part accounting for all irreversible changes. The second concept---the maximization of the rate of entropy production---determines the constitutive equation for the Cauchy stress tensor and specifies the evolution equation for the elastic part of the Cauchy stress tensor based solely on constitutive relations for two scalar quantities describing how energy is stored in the body and how it is dissipated.

A key advantage of this framework is that standard viscoelastic rate-type models can be derived without introducing ad hoc kinematic assumptions beyond the specification of two scalar constitutive ingredients, namely the Helmholtz free energy and the rate of dissipation. By postulating the rate of dissipation to be non-negative, compatibility with the second law of thermodynamics is automatically ensured.

Within their approach to viscoelasticity, Rajagopal and Srinivasa \cite{RajagopalSrinivasa2000} focused on incompressible fluids (i.e.~all admissible motions are isochoric), for which the finite elastic response between the underlying natural configuration and the current configuration is also assumed to be isochoric. The Oldroyd-B model is then obtained only after linearization of the elastic response, corresponding to a small-strain approximation. This procedure gives the impression that the Oldroyd-B model is not a fully non-linear constitutive model but rather an approximation arising from a linearized elastic response. By not restricting the elastic response between the natural and current configurations to be isochoric, while maintaining the constraint that the overall motion of the body is incompressible, Málek, Rajagopal, and Tůma \cite{MaRaTu2015} showed that thermodynamic consistency of the Oldroyd-B model can be achieved without invoking a small-strain approximation. 

The objective time derivative appearing in the constitutive equation for the elastic part of the Cauchy stress tensor is not postulated within the thermodynamic framework of Rajagopal and Srinivasa \cite{RajagopalSrinivasa2000}, but instead emerges naturally from the thermodynamic derivation. In this case, the resulting objective derivative is the upper-convected Oldroyd derivative. The methodology for incorporating the Gordon--Schowalter derivative into the thermodynamic framework was subsequently developed by Dostalík, Pr\r{u}\v{s}a and Sk\v{r}ivan \cite{DostalikPrusaSkrivan2019}; see below for further details.

In this study, we extend the Oldroyd-B model with the Gordon--Schowalter time derivative so as to obtain a formulation that generalizes both the Giesekus and the Oldroyd-B models with this objective derivative.
We call such type of fluids Johnson--Segalman--Giesekus (JSG) models. We shall consider two types of such models:
\begin{description}
\item[Model I:] A thermodynamically consistent JSG model presented in \Cref{sec:JSG_TD}, inspired by the derivation in \cite{DostalikPrusaSkrivan2019}.
\item[Model II:] The commonly used engineering variant of the JSG model presented in \Cref{sec:JSG_TD}.
\end{description}

A thermodynamic derivation of a particular variant of the Giesekus model within the Rajagopal–Srinivasa framework was previously presented in \cite{MaRaTu2018}, while its stability analysis and numerical simulations were investigated in \cite{DostalikPrusaTuma2019}.
The motivation for including the Giesekus term into the model is also mathematical: when the higher dissipative term characteristic of the Giesekus model is present, the associated initial–boundary value problems posed on bounded domains with no-slip velocity boundary conditions are known to admit global-in-time weak solutions; see \cite{Masmoudi2011, Los1, Los2}, whereas the corresponding question remains open for the Oldroyd-B model.

Besides the thermodynamic analysis of both models, we assess their performance in the rod climbing configuration and compare the results with experimental measurements \cite{Beavers_Joseph_1975,DEBBAUT1992103}, earlier numerical studies based on the standard engineering Johnson--Segalman and Giesekus models \cite{FIGUEIREDO201698,LUO1999393}, as well as with a recent analytical solution for the Giesekus model \cite{Ruangkriengsin_Brandão_Wu_Hwang_Boyko_Stone_2025}.

Based on the state of the art described above, this article is organized as follows.
In \Cref{sec:JSG_TD}, we introduce Model I and II and show that unlike Model I, the commonly used engineering variant Model~II fails to comply with the second law of
thermodynamics. In \Cref{sec:JSG_deriv}, we derive Model~I
within the Rajagopal--Srinivasa framework in a particular case with the upper convected Oldroyd derivative. In
\Cref{sec:rod_climbing_decs}, we summarize the rod climbing effect and the
associated physical mechanisms. Since rod climbing is a free-surface
phenomenon, a numerical approach capable of handling moving boundaries is
required. The corresponding weak formulation and its Arbitrary
Lagrangian--Eulerian (ALE) counterpart are presented in \Cref{sec:ALE}.
\Cref{sec:numerics} details the numerical implementation. Finally, in
\Cref{sec:results}, we report the results and show that Model~I fits the
experimental data substantially better than the previously reported modeling
attempts.

\section{Two Johnson--Segalman--Giesekus models and their thermodynamic properties}\label{sec:JSG_TD}

The setting considered in this paper is governed by the equations that describe flows of an incompressible isothermal fluid with constant density $\rho$, the incompressibility constraint and the balance of linear momentum. These take the form:
\begin{equation}
\div \mathbf{v} = 0,
\qquad
\rho \overset{\bullet}{\mathbf{v}} = \div \mathbb{T}.
\label{eq:governing_eq}
\end{equation}
Here, $\mathbf{v}$ denotes the velocity field, $\mathbb{T}$ is the Cauchy
stress tensor, and $\overset{\bullet}{z}:= \partial_t z + (\mathbf{v}\cdot\nabla) z$ denotes the material (or convective) derivative, which acts naturally on scalar fields $z$, vector fields $\mathbf{z}$, or tensor fields $\mathbb{Z}$. The Rajagopal and Srinivasa thermodynamic approach \cite{RajagopalSrinivasa2000, RajagopalSrinivasa2004}, recalled in \Cref{sec:JSG_deriv} for clarity of this study, provides the  constitutive relation for $\mathbb{T}$ as well as the evolutionary tensorial equation for the elastic part of the Cauchy stress from the knowledge of two constitutive equations for the Helmholtz free energy and for the rate of dissipation, thereby closing \eqref{eq:governing_eq} in a way that is compatible with the second law.

In this section, we provide mathematical formulations of Model~I and Model~II, as described in the Introduction. Both variants employ the same kinematics, the
same Gordon--Schowalter objective derivative parameterized by $a\in[-1,1]$, and
the same interpretation of the conformation tensor $\mathbb{B}$ as a measure of macroscopic 
elastic deformation in the polymer network\footnote{Note that $\mathbb{B}$ is not the classical left Cauchy--Green tensor.}.
They differ only in the structure
of the Cauchy stress tensor and in the right-hand side of the evolution equation
for $\mathbb{B}$. 
For a tensor field $\mathbb{A}$, the Gordon--Schowalter derivative
is defined as
\begin{equation}
\overset{\triangledown_a}{\mathbb{A}}
=
\overset{\bullet}{\mathbb{A}}
-
a\bigl(\mathbb{D}\mathbb{A}+\mathbb{A}\mathbb{D}\bigr)
-
\bigl(\mathbb{W}\mathbb{A}-\mathbb{A}\mathbb{W}\bigr),
\label{eq:GSder}
\end{equation}
where $\mathbb{L}=\nabla\mathbf{v}$,
$\mathbb{D}=\tfrac12(\mathbb{L}+\mathbb{L}^{\mathsf T})$, and
$\mathbb{W}=\tfrac12(\mathbb{L}-\mathbb{L}^{\mathsf T})$.
For $a=1$, \eqref{eq:GSder} reduces to the upper-convected Oldroyd derivative (Oldroyd-B model), whereas for $a=-1$ it reduces to the lower-convected Oldroyd derivative (Oldroyd-A model).

\medskip

\noindent\textbf{Model~I: A thermodynamically consistent JSG model.}
This formulation extends the conformation-tensor-based Johnson--Segalman model of
Dostalík, Průša and Skřivan \cite{DostalikPrusaSkrivan2019} by incorporating a
Giesekus-type nonlinear relaxation term, and reduces to that model for
$\alpha=0$ and $a\in[-1,1]$. If $\alpha=1$ and
$a=1$, the model recovers a particular variant of the Giesekus model whose
thermodynamic derivation was presented in \cite{MaRaTu2018}. The intermediate case
$\alpha=1/2$ with $a=1$ was considered numerically in
\cite{DostalikPrusaTuma2019}.
The constitutive equations of Model~I read
\begin{equation}
\boxed{
\begin{aligned}
\mathbb{T} = -p\mathbb{I} + 2\mu_s\,\mathbb{D} + aG(\mathbb{B}-\mathbb{I}),&
\\[0.2em]
\overset{\triangledown_a}{\mathbb{B}}
+ \frac{1}{\tau}\left((1-\alpha)(\mathbb{B}-\mathbb{I})
+ \alpha(\mathbb{B}^2-\mathbb{B})\right) &= \mathbb{O}.
\end{aligned}
}
\label{eq:modelA}
\end{equation}
Here, $p$ denotes the pressure, $\mu_s$ the Newtonian solvent viscosity,
$\mathbb{B}$ the conformation tensor, $G>0$ the elastic modulus, $\tau>0$ the
relaxation time, $\mathbb{I}$ the identity tensor and $\mathbb{O}$ the zero
tensor. Note that for $a=0$,
corresponding to the corotational Jaumann--Zaremba derivative, the polymeric contribution to the Cauchy stress vanishes and the velocity evolves according to the classical Navier--Stokes equations.

\medskip

\noindent\textbf{Model~II: Engineering JSG model.}
We consider the classical engineering Johnson--Segalman model formulated in
terms of the polymeric extra stress $\mathbb{S}$
\cite{JohnsonSegalman1977,BirdArmstrongHassager1987,RaoRajagopal1999} and extend it by a
Giesekus-type non-linear relaxation term.
The stress-based formulation reads
\begin{equation}
\begin{aligned}
\mathbb{T} = -p\mathbb{I} + 2\mu_s\,\mathbb{D} + \mathbb{S},
\\[0.2em]
\overset{\triangledown_a}{\mathbb{S}}
+ \frac{1}{\tau}\left(\mathbb{S} + \frac{\alpha}{G}\mathbb{S}^2\right)
= \frac{2\mu_p}{\tau}&\,\mathbb{D},
\end{aligned}
\label{eq:modelB-S}
\end{equation}
where $\mu_p>0$ is the polymer viscosity and $\tau>0$ the relaxation time. For
$\alpha=0$, the system reduces to the classical engineering Johnson--Segalman
model for an arbitrary $a\in[-1,1]$.

Introducing the conformation tensor $\mathbb{B}$ via
\begin{equation}\label{eq:GS-I}
\mathbb{S} = G(\mathbb{B}-\mathbb{I}),
\qquad
\mu_p = \tau G,
\end{equation}
the model can be equivalently rewritten in conformation-tensor form as
\begin{equation}
\boxed{
\begin{aligned}
\mathbb{T} = -p\mathbb{I} + 2\mu_s\,\mathbb{D} + G(\mathbb{B}-\mathbb{I}),
\\[0.2em]
\overset{\triangledown_a}{\mathbb{B}}
+ \frac{1}{\tau}\left((1-\alpha)(\mathbb{B}-\mathbb{I})
+ \alpha(\mathbb{B}^2-\mathbb{B})\right)
= 2&(1-a)\,\mathbb{D}.
\end{aligned}
}
\label{eq:modelB}
\end{equation}
As shown in the subsequent section, when $a\neq 1$, this engineering JSG formulation does not,
in general, admit a positive definite rate of dissipation. 

\subsection{Thermodynamic consistency of Model~I}\label{model:A}

We refer to \cite{MaRaTu2018} for more details regarding the nomenclature in this and the following sections.
We employ a neo-Hookean Helmholtz free energy $\psi$ per unit mass that depends only on
the conformation tensor $\mathbb{B}$, which is assumed to satisfy the second equation in \eqref{eq:modelA}, in the following manner:
\begin{equation}
\psi(\mathbb{B})
=
\frac{G}{2\rho}
\bigl(
\mathrm{tr}\,\mathbb{B}
-
\ln\det\mathbb{B}
-
d
\bigr),
\end{equation}
where $d$ denotes the spatial dimension. The derivative of $\psi$ with respect
to $\mathbb{B}$ is given by
\begin{equation}
\frac{\partial\psi}{\partial\mathbb{B}}
=
\frac{G}{2\rho}(\mathbb{I}-\mathbb{B}^{-1}).
\end{equation}
Since the density $\rho$ is constant and $\psi$ depends only on $\mathbb{B}$,
we obtain
\begin{equation}\label{pepa1}
\rho\overset{\bullet}{\psi}
=
\frac{G}{2}(\mathbb{I}-\mathbb{B}^{-1}):\overset{\bullet}{\mathbb{B}}.
\end{equation}
In the isothermal setting, the reduced thermodynamic identity involving the rate of dissipation $\xi$ takes the form
\begin{equation}
\xi
=
\mathbb{T}:\mathbb{D}
-
\rho\overset{\bullet}{\psi} \quad \text{ with } \quad 
\xi \ge 0.
\end{equation}
We now express $\xi$ using
the constitutive form of the Cauchy stress tensor $\mathbb{T}$ (the first equation in \eqref{eq:modelA}), \eqref{pepa1} and the the evolution equation for $\mathbb{B}$ (the first equation in \eqref{eq:modelA}). Specifically, 
as the Cauchy stress tensor is given by
\begin{equation}
\mathbb{T}
=
-p\mathbb{I}
+
2\mu_s\,\mathbb{D}
+
aG(\mathbb{B}-\mathbb{I}),
\end{equation}
using the incompressibility ($\operatorname{tr}\mathbb{D}=0$), we observe that the stress power takes the form
\begin{equation}
\mathbb{T}:\mathbb{D}
=
2\mu_s|\mathbb{D}|^2
+
aG\,\mathbb{B}:\mathbb{D}.
\end{equation}
Next, we express $\overset{\bullet}{\mathbb{B}}$ in terms of the Gordon--Schowalter derivative (see \eqref{eq:GSder}):
\begin{equation}\label{eq:Bdot_using_Ba}
\overset{\bullet}{\mathbb{B}}
=
\overset{\triangledown_a}{\mathbb{B}}
+
a(\mathbb{D}\mathbb{B}+\mathbb{B}\mathbb{D})
+
(\mathbb{W}\mathbb{B}-\mathbb{B}\mathbb{W}).
\end{equation}
Then, it follows from \eqref{pepa1} that, for symmetric positive definite $\mathbb{B}$,
\begin{equation}
\rho\overset{\bullet}{\psi}
=
\frac{G}{2}(\mathbb{I}-\mathbb{B}^{-1})
:\overset{\triangledown_a}{\mathbb{B}}
+
aG\,\mathbb{B}:\mathbb{D}.
\end{equation}
Finally, using the evolution equation of Model~I given by \eqref{eq:modelA}, we obtain that 
\begin{equation}
\rho\overset{\bullet}{\psi}
=
-\frac{G}{2\tau}
(\mathbb{I}-\mathbb{B}^{-1})
:
\left(
(1-\alpha)(\mathbb{B}-\mathbb{I})
+
\alpha(\mathbb{B}^2-\mathbb{B})
\right)
+
aG\,\mathbb{B}:\mathbb{D}.
\end{equation}

\subsubsection{Rate of dissipation and its positivity}

The rate of dissipation therefore reads
\begin{align}
\xi
&=
\mathbb{T}:\mathbb{D}
-
\rho\overset{\bullet}{\psi}
\\
&=
2\mu_s|\mathbb{D}|^2
+
\frac{G}{2\tau}
(\mathbb{I}-\mathbb{B}^{-1})
:
\left(
(1-\alpha)(\mathbb{B}-\mathbb{I})
+
\alpha(\mathbb{B}^2-\mathbb{B})
\right)\\
&=
2\mu_s|\mathbb{D}|^2
+
(1-\alpha)\frac{G}{2\tau}
(\mathbb{I}-\mathbb{B}^{-1})
:
(\mathbb{B}-\mathbb{I})
+
\alpha\frac{G}{2\tau}|\mathbb{B}-\mathbb{I}|^2.
\end{align}
Diagonalising the symmetric positive definite tensor $\mathbb{B}$ as
\begin{equation}
\mathbb{B}
=
\mathbb{Q}\,\mathrm{diag}(\lambda_1,\dots,\lambda_d)\,\mathbb{Q}^{\mathsf{T}},
\qquad
\lambda_i>0,
\end{equation}
we obtain
\begin{equation}
(\mathbb{I}-\mathbb{B}^{-1}):(\mathbb{B}-\mathbb{I})
=
\sum_{i=1}^d
(\lambda_i^{1/2}-\lambda_i^{-1/2})^2
\ge 0.
\end{equation}
The remaining term $|\mathbb{B}-\mathbb{I}|^2$ is manifestly non-negative.
Consequently, the rate of dissipation satisfies
\begin{equation}
\boxed{
\xi
=
2\mu_s|\mathbb{D}|^2
+
(1-\alpha)\frac{G}{2\tau}
(\mathbb{I}-\mathbb{B}^{-1})
:
(\mathbb{B}-\mathbb{I})
+
\alpha\frac{G}{2\tau}|\mathbb{B}-\mathbb{I}|^2
\ge 0.
}
\end{equation}

Model~I is therefore thermodynamically consistent for all $a\in[-1,1]$ and
$\alpha\in[0,1]$. Note that the rate of dissipation $\xi$ does not depend on the
parameter $a$; consequently, Model~I satisfies the second law of thermodynamics
for all $a\in\mathbb{R}$.

\subsection{Thermodynamic analysis of Model~II}\label{model:B}

We now turn to Model~II. This model employs the same Helmholtz free energy
$\psi(\mathbb{B})$ as Model~I, but the Cauchy stress tensor takes a different form, in
which the Gordon--Schowalter parameter $a$ does not appear, namely 
\begin{equation}
\mathbb{T}
=
-p\mathbb{I}
+
2\mu_s\,\mathbb{D}
+
G(\mathbb{B}-\mathbb{I}).
\end{equation}
The corresponding stress power contribution reads
\begin{equation}
\rho\overset{\bullet}{\psi}
=
\frac{G}{2}(\mathbb{I}-\mathbb{B}^{-1}):\overset{\bullet}{\mathbb{B}}.
\end{equation}
As before, we express $\rho\overset{\bullet}{\psi}$ by substituting for $\overset{\bullet}{\mathbb{B}}$
from \eqref{eq:Bdot_using_Ba},
\begin{equation}
\rho\overset{\bullet}{\psi}
=
\frac{G}{2}(\mathbb{I}-\mathbb{B}^{-1})
:\overset{\triangledown_a}{\mathbb{B}}
+
a\,G\,\mathbb{B}:\mathbb{D}.
\end{equation}
Substituting the evolution law of Model~II given in \eqref{eq:modelB}, we obtain
\begin{equation}
\rho\overset{\bullet}{\psi}
=
-\frac{G}{2\tau}
(\mathbb{I}-\mathbb{B}^{-1})
:
\left(
(1-\alpha)(\mathbb{B}-\mathbb{I})
+
\alpha(\mathbb{B}^2-\mathbb{B})
\right)
+
G\bigl(
a\,\mathbb{B}:\mathbb{D}
-
(1-a)\mathbb{B}^{-1}:\mathbb{D}
\bigr).
\end{equation}

\subsubsection{Rate of dissipation}

The rate of dissipation $\xi=\mathbb{T}:\mathbb{D}-\rho\overset{\bullet}{\psi}$ therefore reads
\begin{align}
\xi
&=
2\mu_s|\mathbb{D}|^2
+
\frac{G}{2\tau}
(\mathbb{I}-\mathbb{B}^{-1})
:
\left(
(1-\alpha)(\mathbb{B}-\mathbb{I})
+
\alpha(\mathbb{B}^2-\mathbb{B})
\right)
\\
&\quad
+
G(1-a)\bigl(\mathbb{B}+\mathbb{B}^{-1}\bigr):\mathbb{D}\\
&=
2\mu_s|\mathbb{D}|^2
+
(1-\alpha)\frac{G}{2\tau}
(\mathbb{I}-\mathbb{B}^{-1})
:
(\mathbb{B}-\mathbb{I})
+
\alpha\frac{G}{2\tau}|\mathbb{B}-\mathbb{I}|^2
\\
&\quad
+
G(1-a)\bigl(\mathbb{B}+\mathbb{B}^{-1}\bigr):\mathbb{D}.
\end{align}
Thus,
\begin{equation}
\boxed{
\xi
=
2\mu_s|\mathbb{D}|^2
+
(1-\alpha)\frac{G}{2\tau}
(\mathbb{I}-\mathbb{B}^{-1})
:
(\mathbb{B}-\mathbb{I})
+
\alpha\frac{G}{2\tau}|\mathbb{B}-\mathbb{I}|^2
+
G(1-a)\bigl(\mathbb{B}+\mathbb{B}^{-1}\bigr):\mathbb{D}.
}
\end{equation}
The first three terms are non-negative, as in Model~I.  
The difficulty arises from the mixed term
\begin{equation}
G(1-a)\bigl(\mathbb{B}+\mathbb{B}^{-1}\bigr):\mathbb{D},
\end{equation}
which has no definite sign. In particular, one can choose $\mathbb{B}$ and
$\mathbb{D}$ such that this term becomes negative and dominates the
non-negative contributions.

\medskip

In contrast to Model~I, the engineering Johnson--Segalman model in the
conformation-tensor form \eqref{eq:modelB} does \emph{not} guarantee
non-negative rate of dissipation for $a\neq 1$. In the special case $a=1$
(the upper-convected Oldroyd derivative) and for arbitrary $\alpha\in[0,1]$ do
Model~I and Model~II coincide, and the rate of dissipation reduces to a positive
definite form. This observation explains why thermodynamically consistent
Johnson--Segalman-type models, such as those proposed in
\cite{DostalikPrusaSkrivan2019}, must differ from the classical engineering
formulation.

\begin{remark}[Remarks on the Johnson--Segalman Model~II in the rheological literature]
The Johnson--Segalman model was originally introduced as a phenomenological generalization of the Oldroyd-B model aimed at capturing non-affine deformation effects and non-monotonic shear stress responses in complex fluids. Its primary motivation was to provide a simple constitutive framework capable of reproducing experimentally observed shear-thinning behavior and shear banding in steady homogeneous shear flows \cite{JohnsonSegalman1977}. As such, the model has been widely employed as an engineering interpolation rather than as a constitutive equation derived from microstructural or energetic considerations.

One manifestation of this limitation is evident in the model's response to time-dependent flows. In \cite[Chapter 4]{huilgol1997fluid}, Huilgol and Phan-Thien analytically computed the Johnson--Segalman response in a single-step strain and showed that it does not obey the Lodge--Meissner rule. This example illustrates that, while the model captures certain shear flow phenomena, it can produce stress responses, particularly in flows dominated by elastic effects, that cannot be fully reconciled with a physically meaningful elastic energy interpretation.
\end{remark}

\section{Thermodynamically consistent derivation of the Johnson--Segalman--Giesekus model}\label{sec:JSG_deriv}

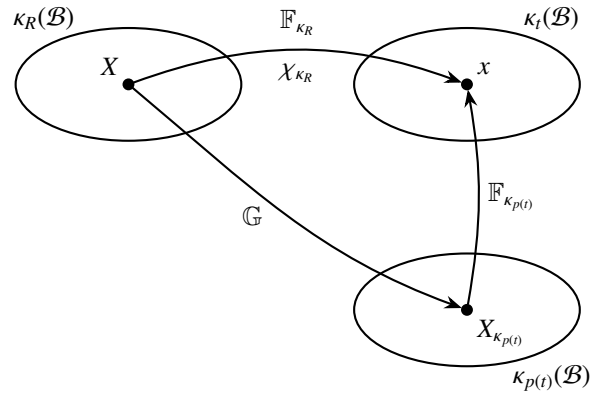
\begin{figure}[h!]
    \centering
\begin{tikzpicture}[scale=1.5, every node/.style={scale=1},  >=Stealth, thick]

\draw (0,1) ellipse (1cm and 0.5cm); %
\draw (3,1) ellipse (1cm and 0.5cm); %
\draw (3,-1) ellipse (1cm and 0.5cm); %

\fill (0,1) circle (1.5pt); 
\node[above left] at (0,1) {$X$};

\fill (3,1) circle (1.5pt); 
\node[above right] at (3,1) {$x$};

\fill (3,-1) circle (1.5pt); 
\node[below right] at (3,-1) {$X_{\kappa_{p(t)}}$};

\node[above] at (-0.75,1.4) {$\kappa_R(\mathcal{B})$};
\node[above] at (3.75,1.4) {$\kappa_t(\mathcal{B})$};
\node[below] at (3.75,-1.4) {$\kappa_{p(t)}(\mathcal{B})$};

\draw[->, shorten >=2pt, shorten <=2pt]
  (0,1) to[out=20, in=160] (3,1);        %

\draw[->, shorten >=2pt, shorten <=2pt]
  (0,1) to[out=-40, in=160] (3,-1);      %

\draw[->, shorten >=2pt, shorten <=2pt]
  (3,-1) to[out=80, in=-80] (3,1);       %

\node[above] at (1.5,1.35) {$\mathbb{F}_{\kappa_R}$}; %
\node[below] at (1.5,1.27) {$\chi_{\kappa_R}$};      %
\node[below] at (1.1,0) {$\mathbb{G}$};        %
\node[right] at (3.1,0) {$\Fp$};            %

\end{tikzpicture}

    \caption{Sketch of the reference configuration $\kappa_R(\mathcal{B})$, the current configuration $\kappa_t(\mathcal{B})$ and the natural configuration $\kappa_{p(t)}(\mathcal{B})$. The deformation gradient $\mathbb{F}_{\kappa_R}$ is multiplicatively decomposed. The natural configuration is defined as the configuration that the body in the current configuration would take if the external stimuli were removed. Hence the natural configuration $\kappa_{p(t)}(\mathcal{B})$ is associated with the current configuration $\kappa_t(\mathcal{B})$ and it evolves with the body as the body produces entropy. This allows us to split the total deformation $\mathbb{F}_{\kappa_R}$ into a purely elastic (reversible) part $\Fp : \kappa_{p(t)}(\mathcal{B}) \longrightarrow \kappa_{t}(\mathcal{B})$ and the rest (dissipation) $\mathbb{G} : \kappa_{R}(\mathcal{B}) \longrightarrow \kappa_{p(t)}(\mathcal{B})$ so that: $\mathbb{F}_{\kappa_R} = \Fp \mathbb{G}$.}
    \label{fig:natural_config}
\end{figure}

In this section we present a thermodynamically consistent derivation of the
Johnson--Segalman--Giesekus (JSG) model restricted to the case $a=1$,
within the thermodynamic framework based on evolving natural configuration,
see \Cref{fig:natural_config}.
The constitutive theory is formulated by prescribing the Helmholtz free energy
and the rate  of entropy production as the fundamental constitutive ingredients.
The Cauchy stress tensor and the evolution equation for the elastic conformation tensor
are then identified from the reduced thermodynamic identity.
A thorough overview of this derivation framework, including an explanation of the thermodynamic background, can be found in~\cite{MaPr2018}.

\subsection{Kinematics and natural configuration}

The constitutive framework employed in this section is based on the notion of a
\emph{time-evolving natural configuration}, understood as an evolving state of the
material associated with micro-structural processes and entropy production
\cite{RajagopalSrinivasa2000}.
The natural configuration provides a kinematic reference that enables a
decomposition of the total deformation into an elastic, energy-storing part and
a dissipative part.
We therefore employ the multiplicative decomposition
\begin{equation}
\mathbb{F} = \mathbb{F}_{\kappa_{p(t)}} \mathbb{G},
\end{equation}
where $\mathbb{F}_{\kappa_{p(t)}}$ maps from the natural configuration
$\kappa_{p(t)}(\mathcal{B})$ to the current configuration $\kappa_t(\mathcal{B})$, while $\mathbb{G}$ maps the
reference configuration $\kappa_R(\mathcal{B})$ to $\kappa_{p(t)}(\mathcal{B})$.
Within this framework, $\mathbb{F}_{\kappa_{p(t)}}$ is responsible for the elastic
part of the deformation, whereas $\mathbb{G}$ accounts for processes associated with entropy production.
In what follows, we assume that the total deformation is incompressible, that is
$\det \mathbb{F} = 1$. However, in line with the approach adopted in
\cite{MaRaTu2015}, no incompressibility constraint is imposed on the individual
mappings $\mathbb{F}_{\kappa_{p(t)}}$ and $\mathbb{G}$, which are therefore allowed
to be compressible.

The left Cauchy--Green tensor (symmetric positive definite) associated with the natural configuration
is defined as
\begin{equation}
\mathbb{B}_{\kappa_{p(t)}}
:=
\mathbb{F}_{\kappa_{p(t)}} \mathbb{F}_{\kappa_{p(t)}}^{\mathsf T},
\end{equation}
while the right Cauchy--Green tensor (symmetric positive definite) is defined as
\begin{equation}
\mathbb{C}_{\kappa_{p(t)}}
:=
\mathbb{F}_{\kappa_{p(t)}}^{\mathsf T} \mathbb{F}_{\kappa_{p(t)}}.
\end{equation}
The tensors $\mathbb{B}_{\kappa_{p(t)}}$ and $\mathbb{C}_{\kappa_{p(t)}}$ are similar
and therefore share the same eigenvalues.
The evolution of the
natural configuration is described by
\begin{equation}
\mathbb{L}_{\kappa_{p(t)}}
:=
\overset{\bullet}{\mathbb{G}}\,\mathbb{G}^{-1},
\qquad
\mathbb{D}_{\kappa_{p(t)}}
:=
\tfrac12\bigl(\mathbb{L}_{\kappa_{p(t)}}+\mathbb{L}_{\kappa_{p(t)}}^{\mathsf T}\bigr).
\end{equation}
Differentiating $\mathbb{F}=\mathbb{F}_{\kappa_{p(t)}}\mathbb{G}$ yields
\begin{equation}
\overset{\bullet}{\mathbb{F}}_{\kappa_{p(t)}}
=
\mathbb{L}\mathbb{F}_{\kappa_{p(t)}}
-
\mathbb{F}_{\kappa_{p(t)}}\mathbb{L}_{\kappa_{p(t)}}.
\end{equation}
Consequently, the upper-convected Oldroyd derivative of
$\mathbb{B}_{\kappa_{p(t)}}$ satisfies the exact kinematic identity
\begin{equation}
\overset{\triangledown}{\mathbb{B}}_{\kappa_{p(t)}}
:=
\overset{\bullet}{\mathbb{B}}_{\kappa_{p(t)}}
-
\mathbb{L}\mathbb{B}_{\kappa_{p(t)}}
-
\mathbb{B}_{\kappa_{p(t)}}\mathbb{L}^{\mathsf T}
=
-2\,\mathbb{F}_{\kappa_{p(t)}}\mathbb{D}_{\kappa_{p(t)}}
\mathbb{F}_{\kappa_{p(t)}}^{\mathsf T}.
\label{eq:ucB_nat}
\end{equation}

\subsection{Energetic structure and thermodynamics}

\paragraph{Free energy}

We prescribe the neo-Hookean Helmholtz free energy per unit mass as a function of
the elastic conformation tensor, 
\begin{equation}
\psi(\mathbb{B}_{\kappa_{p(t)}})
=
\frac{G}{2\rho}
\bigl(
\operatorname{tr}\mathbb{B}_{\kappa_{p(t)}}
-
\ln\det\mathbb{B}_{\kappa_{p(t)}}
-
d
\bigr),
\end{equation}
which represents an isotropic elastic response associated solely with the
deformation between the natural and the current configurations.
Since the density $\rho$ is constant, the material time derivative of the free
energy reads
\begin{equation}
\rho\overset{\bullet}{\psi}
=
\frac{G}{2}
\bigl(\mathbb{I}-\mathbb{B}_{\kappa_{p(t)}}^{-1}\bigr)
:
\overset{\triangledown}{\mathbb{B}}_{\kappa_{p(t)}}
+
G\,\mathbb{B}_{\kappa_{p(t)}}:\mathbb{D}.
\label{eq:psidot_nat}
\end{equation}

\paragraph{Reduced thermodynamic identity}

Following the thermodynamic framework recalled in \Cref{sec:JSG_TD}, we employ the
reduced thermodynamic identity
\begin{equation}
\xi = \mathbb{T}:\mathbb{D} - \rho\overset{\bullet}{\psi},
\end{equation}
which links the stress power, the rate of change of the Helmholtz free energy,
and the rate of dissipation.
Substituting \eqref{eq:psidot_nat} yields
\begin{equation}
\xi
=
\bigl(\mathbb{T}-G\mathbb{B}_{\kappa_{p(t)}}\bigr):\mathbb{D}
-
\frac{G}{2}
\bigl(\mathbb{I}-\mathbb{B}_{\kappa_{p(t)}}^{-1}\bigr)
:
\overset{\triangledown}{\mathbb{B}}_{\kappa_{p(t)}}.
\end{equation}
Using \eqref{eq:ucB_nat} 
we obtain
\begin{equation}
-\frac{G}{2}
\bigl(\mathbb{I}-\mathbb{B}_{\kappa_{p(t)}}^{-1}\bigr)
:
\overset{\triangledown}{\mathbb{B}}_{\kappa_{p(t)}}
=
G(\mathbb{C}_{\kappa_{p(t)}}-\mathbb{I})
:
\mathbb{D}_{\kappa_{p(t)}}.
\end{equation}
Hence, the rate of dissipation computed from the Helmholtz free energy reads
\begin{equation}
\boxed{
\xi
=
\bigl(\mathbb{T}^d-G\mathbb{B}_{\kappa_{p(t)}}^{\mathrm d}\bigr):\mathbb{D}
+
G(\mathbb{C}_{\kappa_{p(t)}}-\mathbb{I})
:
\mathbb{D}_{\kappa_{p(t)}},
}
\label{eq:xi_split_nat}
\end{equation}
where $\mathbb{B}_{\kappa_{p(t)}}^{\mathrm d}$ denotes the deviatoric part of
$\mathbb{B}_{\kappa_{p(t)}}$.

\subsection{Rate of dissipation and constitutive structure}
\paragraph{Prescribed rate of dissipation}

Guided by the structure of \eqref{eq:xi_split_nat}, we prescribe the rate of dissipation in the form
\begin{equation}\label{eq:xi_pres_correct}
\boxed{
\xi
=
2\mu_s|\mathbb{D}|^2
+
2\mu_p\,
\mathbb{D}_{\kappa_{p(t)}}
\mathbb{M}(\mathbb{C}_{\kappa_{p(t)}})
:
\mathbb{D}_{\kappa_{p(t)}},
}
\end{equation}
where $\mu_s\geq 0$, $\mu_p\geq 0$, and the mobility tensor is defined as
\begin{equation}
\mathbb{M}(\mathbb{C}_{\kappa_{p(t)}})
=
\bigl((1-\alpha)\mathbb{C}_{\kappa_{p(t)}}^{-1}
+
\alpha\mathbb{I}\bigr)^{-1},
\qquad
\alpha\in[0,1].
\end{equation}
For $\alpha=0$ this choice reduces to the classical Oldroyd-B rate of dissipation 
$2\mu_p\,\mathbb{D}_{\kappa_{p(t)}}\mathbb{C}_{\kappa_{p(t)}}:\mathbb{D}_{\kappa_{p(t)}}$,
while for $\alpha=1$ it yields $2\mu_p|\mathbb{D}_{\kappa_{p(t)}}|^2$, corresponding to the simplified Giesekus derived in \cite{MaRaTu2018}. Since the mobility tensor $\mathbb{M}(\mathbb{C}_{\kappa_{p(t)}})$ is symmetric positive definite, being the inverse of a convex combination of two symmetric positive definite matrices, the rate of dissipation is non-negative, $\xi\ge0$, and the second law of thermodynamics is satisfied.

\paragraph{Principle of maximal rate of entropy production}

At this stage, the reduced thermodynamic identity
\eqref{eq:xi_split_nat} and the prescribed form of the rate of dissipation
\eqref{eq:xi_pres_correct} provide two scalar expressions for the rate of
entropy production $\xi$.
Equating these expressions alone is, however, not sufficient to uniquely
determine the constitutive relations, since infinitely many tensorial
relations between $\mathbb{D}_{\kappa_{p(t)}}$ and $\mathbb{C}_{\kappa_{p(t)}}$
may lead to the same scalar value of $\xi$.

The crucial additional constitutive principle, introduced by Rajagopal and
Srinivasa \cite{RajagopalSrinivasa2000}, is the
\emph{principle of maximal rate of entropy production}.
According to this principle, the actual evolution of the natural configuration
is selected among all kinematically admissible processes so as to maximize the
rate of entropy production, subject to the constraint
$\xi=\mathbb{T}:\mathbb{D}-\rho\overset{\bullet}{\psi}$ imposed by the reduced thermodynamic
identity.
In this way, the principle acts as a selection mechanism: it resolves the
non-uniqueness inherent in equating scalar rates of entropy production and renders the
underlying tensorial constitutive relations unique.

\paragraph{Identification of the constitutive relations}
In the present setting, the \emph{principle of maximal rate of entropy
production} leads to a unique relation between
$\mathbb{D}_{\kappa_{p(t)}}$ and $\mathbb{C}_{\kappa_{p(t)}}$.
Rather than formulating the corresponding constrained maximization problem
explicitly, we adopt an equivalent identification procedure based on
\eqref{eq:xi_split_nat} and \eqref{eq:xi_pres_correct}, which yields the same
constitutive equations while preserving transparency of the derivation, namely
by direct comparison:
\begin{equation}
\mathbb{T}^d = 2\mu_s\D + G\mathbb{B}_{\kappa_{p(t)}}^d
\end{equation}
and the full Cauchy stress tensor $\mathbb{T}=m\mathbb{I}+\mathbb{T}^d$ upon defining $p=-m-G(1-\tr\mathbb{B}_{\kappa_{p(t)}}/3)$ yields
\begin{equation}
\boxed{
\mathbb{T}
=
-p\mathbb{I}
+
2\mu_s\,\mathbb{D}
+
G(\mathbb{B}_{\kappa_{p(t)}}-\mathbb{I}).
}
\label{eq:T_correct}
\end{equation}
When comparing the
terms with $\mathbb{D}_{\kappa_{p(t)}}$ in \eqref{eq:xi_split_nat} and \eqref{eq:xi_pres_correct}, we obtain
\begin{equation}
\label{eq:relDpCp}
G(\mathbb{C}_{\kappa_{p(t)}}-\mathbb{I})
=
2\mu_p\,
\mathbb{D}_{\kappa_{p(t)}}\mathbb{M}(\mathbb{C}_{\kappa_{p(t)}}).
\end{equation}
Multiplying \eqref{eq:relDpCp} by
$\mathbb{F}_{\kappa_{p(t)}}$ from the left
and by $\mathbb{M}(\mathbb{C}_{\kappa_{p(t)}})^{-1}\mathbb{F}_{\kappa_{p(t)}}^{\mathsf T}$ from the right, we obtain 
\begin{equation}
G\left[(1-\alpha)(\mathbb{B}_{\kappa_{p(t)}}-\mathbb{I})
+
\alpha(\mathbb{B}_{\kappa_{p(t)}}^2-\mathbb{B}_{\kappa_{p(t)}})\right]
=
2\mu_p\,
\mathbb{F}_{\kappa_{p(t)}}\mathbb{D}_{\kappa_{p(t)}}
\mathbb{F}_{\kappa_{p(t)}}^{\mathsf T}.
\end{equation}
Introducing $\tau:=\mu_p/G$, we arrive at
\begin{equation}
\boxed{
\overset{\triangledown}{\mathbb{B}}_{\kappa_{p(t)}}
+
\frac{1}{\tau}
\left(
(1-\alpha)(\mathbb{B}_{\kappa_{p(t)}}-\mathbb{I})
+
\alpha(\mathbb{B}_{\kappa_{p(t)}}^2-\mathbb{B}_{\kappa_{p(t)}})
\right)
=
\mathbb{O}.
}
\label{eq:B_evo_correct}
\end{equation}

\subsection{Equivalence with the rate of dissipation written in terms of
$\mathbb{B}_{\kappa_{p(t)}}$}

Substituting $\mathbb{D}_{\kappa_{p(t)}}$ from \eqref{eq:relDpCp} into the
prescribed rate of dissipation \eqref{eq:xi_pres_correct} yields
\begin{equation}
\xi
=
2\mu_p\,
\mathbb{D}_{\kappa_{p(t)}}
:
\mathbb{M}(\mathbb{C}_{\kappa_{p(t)}})
\mathbb{D}_{\kappa_{p(t)}}
=
\frac{G}{2\tau}
\left(
(1-\alpha)
(\mathbb{C}_{\kappa_{p(t)}}-\mathbb{I})
:
\mathbb{C}_{\kappa_{p(t)}}^{-1}
(\mathbb{C}_{\kappa_{p(t)}}-\mathbb{I})
+
\alpha|\mathbb{C}_{\kappa_{p(t)}}-\mathbb{I}|^2
\right).
\end{equation}
Since $\mathbb{B}_{\kappa_{p(t)}}$ and $\mathbb{C}_{\kappa_{p(t)}}$ share eigenvalues,
we have
\begin{equation}
|\mathbb{C}_{\kappa_{p(t)}}-\mathbb{I}|^2
=
|\mathbb{B}_{\kappa_{p(t)}}-\mathbb{I}|^2,
\qquad
(\mathbb{C}_{\kappa_{p(t)}}-\mathbb{I})
:
\mathbb{C}_{\kappa_{p(t)}}^{-1}
(\mathbb{C}_{\kappa_{p(t)}}-\mathbb{I})
=
(\mathbb{I}-\mathbb{B}_{\kappa_{p(t)}}^{-1})
:
(\mathbb{B}_{\kappa_{p(t)}}-\mathbb{I}),
\end{equation}
and therefore the rate of dissipation \eqref{eq:xi_pres_correct} can be equivalently written as 
\begin{equation}
\xi
=
2\mu_s|\mathbb{D}|^2
+
(1-\alpha)\frac{G}{2\tau}
(\mathbb{I}-\mathbb{B}_{\kappa_{p(t)}}^{-1})
:
(\mathbb{B}_{\kappa_{p(t)}}-\mathbb{I})
+
\alpha\frac{G}{2\tau}|\mathbb{B}_{\kappa_{p(t)}}-\mathbb{I}|^2
,
\end{equation}
i.e.\ as the rate of dissipation obtained in the previous section.

\section{The rod climbing (Weissenberg) effect}\label{sec:rod_climbing_decs}
The rod climbing effect, also known as the Weissenberg effect, refers to the rise of a viscoelastic fluid along a rod rotating about its axis. This effect was recently revisited in the experimental study by More et al.~in 2023 \cite{More2023}. In contrast to Newtonian fluids, which typically exhibit a descent of the free surface near the rod due to inertial effects, polymeric fluids may climb the rod as a consequence of elastic normal stresses generated in shear flow.

The flow induced by a rotating rod is predominantly azimuthal and may be viewed locally as a torsional shear flow. For a Newtonian fluid, the shear stress field is fully determined by the shear viscosity and the shear rate, and centrifugal effects lead to a radially outward pressure gradient that depresses the free surface near the rod. In viscoelastic fluids, however, the constitutive response includes normal stress differences. %

In the rotating-rod geometry, the dominant elastic contribution arises from the azimuthal normal stress $\tau_{\theta\theta}$ acting along curved streamlines. This stress may be interpreted as a hoop stress that pulls fluid elements radially inward toward the axis of rotation. As a result, fluid is driven toward the rod, producing excess pressure near the rod surface. In the presence of a deformable free surface, this pressure imbalance is relieved by an upward displacement of the interface, leading to rod climbing.

At steady state, the climbing height is determined by a balance between elastic normal stresses, gravity, surface tension, and inertia. In the limit of low rotation rates, asymptotic analysis based on second-order fluid theory show that the perturbation of the free surface height scales quadratically with the angular velocity of the rod and depends on a specific combination of the normal stress coefficients. In particular, the first normal stress difference promotes climbing, while the second normal stress difference, which is typically negative for polymer solutions, partially counteracts this effect. A recent analytical study \cite{Ruangkriengsin_Brandão_Wu_Hwang_Boyko_Stone_2025} provides explicit conditions for the occurrence of rod climbing of a Giesekus fluid.

Rod climbing thus provides a direct macroscopic manifestation of elastic stresses in shear flow and vanishes in the absence of fluid elasticity. Owing to the sensitivity of the free surface to small stress imbalances, the rotating-rod configuration has long served as a qualitative demonstration of viscoelasticity and, under controlled conditions, as a quantitative probe of normal stress differences at shear rates that are often inaccessible to conventional rheometric techniques. The idea of the rotating rod viscometer comes from \cite{Beavers_Joseph_1975}.

\section{Arbitrary Lagrangian-Eulerian classical and weak formulations}\label{sec:ALE}
To handle problems with moving boundaries or domain, we use the Arbitrary Lagrangian-Eulerian (ALE) method, which is a computational technique that combines elements of both the Lagrangian and the Eulerian approaches. Together with the Lagrangian ($X,\,\Omega_X$) and Eulerian ($x,\,\Omega_x$) configurations, we consider the \textit{mesh} (ALE) configuration ($\chi,\,\Omega_\chi$), which virtually lies between them and there is a mapping from each configuration to the others.  

The ALE method works as follows. In the Eulerian approach, the computational domain is discretized into cells or elements, creating a mesh. This mesh is allowed to move with the motion of the boundaries, allowing deformation and adaptation to the changing shape of the domain. Meanwhile, as the mesh moves, the equations of motion, such as those in the viscoelastic fluid model mentioned earlier, are solved within a Lagrangian reference frame inside the mesh cells. In this context, the fluid properties are monitored as if they are linked to the mobile mesh. The motion of the mesh inside the domain is typically optimized to minimize distortion and ensure favorable mesh quality, which is done within the \textit{mesh} configuration.

We discuss two possible transformations of the model using ALE mappings. The advantage of the Full ALE method is that we solve the balance equations in a fixed computational domain (the \textit{mesh} domain). The price we pay for that is the additional geometric non-linearities that appear in the equations. In some cases, it is possible and more convenient to switch back to the actual configuration but instead of a fully Eulerian approach, evaluate time derivatives at moving mesh grid points. This is the essence of what we would call the Updated ALE method. However, in this paper, we use the Full ALE method only, hereafter referred to as the ALE method. Both these classical approaches were investigated/applied, for instance, in \cite{DONEA1982689, HronTurek2006a,turek2011numerical}; and in the context of viscoelastic fluids \cite{HronRajagopalTuma2014}. For further details, the reader is directed to the monograph of Richter on fluid-structure interaction \cite{Richter2017FSIbook}.

Let us transform the Model~I equations into the \textit{mesh} configuration only to obtain the ALE method. To do this, we introduce a new variable deformation $\ut$ of the mesh that is arbitrary in the domain with the restriction that the deformation is physical on the boundary of the domain. Hence, we identify the mesh with a new configuration $\kappa_\chi$ (domain $\Omega_\chi$), and define the mapping $\phit$ from $\kappa_\chi$ to $\kappa_t$ (domain $\Omega_x$) as $ \phit : \chi \longrightarrow x := \chi + \ut$.

Now we prescribe the mesh deformation $\ut$ in such a way that the material points (i.e.~the time derivative of the displacement equals the velocity) lie only on the boundary $\partial\Omega_\chi$, while inside the domain we allow for an arbitrary but unique solution. Notice that if we enforce the solution inside the domain to also consist of material points, then we recover the Lagrangian formulation. However, we want the solution inside the domain to be as simple as possible, in order to prevent mesh distortion caused by vortical flow. Hence, one may choose, for example, a Laplace equation inside the domain. On the free-surface boundary, we may for simplicity retain the condition that all points are material, and we thus obtain the strong ALE formulation:
\begin{equation}
\begin{split}
-\Delta_{\chi} \ut=&\,0 \quad \text{in} \,\, \Omega_\chi, \\
\frac{\partial \ut}{\partial t} =&\,\vt\quad \text{on} \,\, \partial\Omega_\chi.
\end{split}
\label{03eq02.4}
\end{equation}

Moreover, we define the deformation gradient $\Fb$ and its Jacobian $\Jh = \det\Fb$ as
$\Fb := \frac{\partial \phit}{\partial \chi} = \I+\nabla_{\chi}\ut$. Now, in the same manner as it would be in the case of the Lagrangian formulation, we transform the Eulerian formulation into the ALE formulation. We substitute the velocity gradient:
\begin{equation*}
\nabla_{\chi}\vt = \frac{\partial \vt(t,\phit(t,\chi))}{\partial \chi} = \frac{\partial  \vt(t,x)}{\partial x} \frac{\partial\phit}{\partial \chi} = (\nabla_x \vt)\Fb \,\,\, \Rightarrow \,\,\, \nabla_x \vt = (\nabla_{\chi}\vt)\Fb^{-1},
\end{equation*}
and the material time derivative:
\begin{equation*}
\begin{split}
\frac{\partial\alpha}{\partial t}\Big|_{\chi} &= \frac{d\alpha(\phit(t,\chi),t)}{d t}\Big|_{\chi} = \frac{\partial\alpha}{\partial t}\Big|_x + \frac{\partial \alpha}{\partial x} \frac{\partial\phit}{\partial t}\Big|_{\chi} = \frac{\partial\alpha}{\partial t}\Big|_x + \frac{\partial \ut  }{\partial t} \cdot\nabla_x\alpha \quad \Rightarrow \\
\overset{\bullet}{\alpha} &= \frac{\partial\alpha}{\partial t}\Big|_{\chi} + \left(\vt-\frac{\partial  \ut}{\partial t}\right) \cdot\nabla_x\alpha = \frac{\partial\alpha}{\partial t}\Big|_{\chi} + \left[\Fb^{-1} \left(\vt-\frac{\partial  \ut}{\partial t}\right)\right] \cdot\nabla_{\chi}\alpha.
\end{split}
\end{equation*}
Finally, using the integral substitution theorem from $\Omega_x$ to $\Omega_\chi$ and the consequence of the Piola identity $\rm{div}_\chi\left((\det\Fb)\Fb^{-T}\right) = \mathbf{0}$, we obtain the transformed weak form of Model~I:
\begin{equation}
\begin{split}
\int_{\Omega_\chi} \hat{J}\text{tr}\left((\nabla_{\hat{x}}\mathbf{v})\hat{\mathbb{F}}^{-1}\right)\psi = 0 &, \\
\int_{\Omega_\chi}  \hat{J}\rho\left[\frac{\partial \mathbf{v}}{\partial t} + (\nabla_{\hat{x}}\mathbf{v})\left(\hat{\mathbb{F}}^{-1}(\mathbf{v} - \frac{\partial \hat{\mathbf{u}}}{\partial t}) \right) \right]\cdot\boldsymbol{\phi} + \int_{\Omega_\chi} \hat{J}\hat{\mathbb{T}}\hat{\mathbb{F}}^{-T}\cdot\nabla_{\hat{x}}\boldsymbol{\phi} - \int_{\partial\Omega_\chi} (\hat{J}\hat{\mathbb{T}}\hat{\mathbb{F}}^{-T})\hat{\mathbf{n}}\cdot\boldsymbol{\phi}  - \int_{\Omega_\chi} \hat{J} \rho \mathbf{b} \cdot \boldsymbol{\phi} = 0 &, \\
\hat{\mathbb{T}} = -p\mathbb{I} + 2\mu_s\mathbb{D}_{\hat{x}} + aG(\Bp-\mathbb{I}) &, \\
\int_{\Omega_\chi} \hat{J} \frac{\delta \Bp}{\delta t}\Big|_{\hat{x}} \cdot \mathbb{A} + \hat{J}\frac{1}{\tau}\left[(1-\alpha)(\Bp-\mathbb{I})+ \alpha(\Bp^2-\Bp)\right]\cdot\mathbb{A} =0 &, \\
\frac{\delta \Bp}{\delta t}\Big|_{\hat{x}} = \frac{\partial \Bp}{\partial t}\Big|_{\hat{x}} + (\nabla_{\hat{x}}\Bp)\left(\hat{\mathbb{F}}^{-1}(\mathbf{v} - \frac{\partial \hat{\mathbf{u}}}{\partial t}) \right) - a(\mathbb{D}_{\hat{x}}\Bp + \Bp\mathbb{D}_{\hat{x}}) - (\mathbb{W}_{\hat{x}}\Bp - \Bp\mathbb{W}_{\hat{x}}) &\\
\int_{\Omega_\chi}  \nabla_{\hat{x}} \hat{\mathbf{u}} \cdot \nabla_{\hat{x}} \hat{\mathbf{w}} -\int_{\partial\Omega_\chi} (\nabla\hat{\mathbf{u}})\hat{\mathbf{n}} \cdot \hat{\mathbf{w}} +  \int_{\partial\Omega_\chi} (\nabla\hat{\mathbf{w}})\hat{\mathbf{n}} \cdot \left(\frac{\partial\hat{\mathbf{u}}}{\partial t}-\mathbf{v}\right)+ \frac{\beta_{st}}{h_{min}} \int_{\partial\Omega_\chi} \left(\frac{\partial\hat{\mathbf{u}}}{\partial t}-\mathbf{v}\right) \cdot \hat{\mathbf{w}} = 0&,
\end{split}
\label{weak-nitsche}
\end{equation}
where $2\mathbb{D}_\chi = \left[(\nabla_\chi\vt)\Fb^{-1} + \Fb^{-T} (\nabla_\chi\vt)^T\right] $ and $\hat{\mathbf{n}}$ is a unit outward normal in the \textit{mesh} configuration. This set of equations is closed by the weak formulation of equation~(\ref{03eq02.4}), in which the boundary condition~(\ref{03eq02.4})$_2$ is imposed using the Nitsche method. The parameter $\beta_{st}$ denotes the stabilization (penalty) parameter, and $h_{\min}$ is the minimum element size of the finite-element mesh.

In the following, we introduce a set of simplifications to the governing equations. These simplifications do not alter the physical problem we aim to solve but make the numerical implementation more tractable. The key aspect of combining axi-symmetric cylindrical coordinates $(r,\varphi,z)$, reduced to the meridional plane $(r,z)$, with the ALE method is that the deformation of the mesh in the azimuthal direction is set to zero:
\begin{equation}
    \hat{u}_\varphi = 0.
\end{equation}
This follows from the fact that, under the axi-symmetric assumption, the mesh would otherwise be virtually wrapped around the cylinder. Since the essential role of the ALE method is to preserve the physical shape of the domain boundary, deformation in the azimuthal direction is unnecessary and may be omitted. Consequently, the mesh displacement is taken in the form
\begin{equation}
    \hat{\mathbf{u}} = (\hat{u}_r,\,0,\,\hat{u}_z).
\end{equation}
Furthermore, enforcing the material-point condition on the free surface turns out to be unnecessarily strict, since in the steady state the tangential velocity still enforces tangential motion of the fluid surface. Although physically correct, this choice leads to severe mesh distortion. We therefore leverage the ALE formulation once more and prescribe, on the free-surface boundary, a simple kinematic condition in which the domain deforms only in the $z$-direction. Nevertheless, this condition still accounts for the tangential component of the velocity ($v_r$) and induces the correct surface evolution.

Without loss of generality, we impose this condition in the $(r,z)$ frame rather than in the local frame defined by the actual deformation. This formulation is consistent with the standard kinematic free-surface condition reported in \cite[Eq.~(2.9)]{Ruangkriengsin_Brandão_Wu_Hwang_Boyko_Stone_2025}:
\begin{equation}
\partial_t h = v_z - v_r \partial_r h,
\end{equation}
where, in our setting, the free-surface height is identified as $h := z + \hat{u}_z$, reducing to
\begin{equation}
\hat{\mathbf{u}} = (0,\,0,\,\hat{u}_z)
\end{equation}
only.

In contrast, the velocity field must retain all three components, since it is precisely the azimuthal velocity $v_\varphi$ that drives the secondary flows in the meridional plane. As a result, the weak formulation formally coincides with that of the full three-dimensional problem but is posed on the meridional cross-section $\Omega_\chi$ of the domain. The associated integrals include the standard radial weighting factor $r$, and all spatial derivatives are expressed in cylindrical polar coordinates.

The resulting system of equations for Model~I, including external forces such as gravity and surface tension, is presented below in dimensionless form. For brevity, the previously introduced notation is retained. Model~II can be obtained by straightforward modifications. We arrive at the following complete variational formulation: find  
\begin{equation}
(\mathbf{v},p,\Bp,\hat{\mathbf{u}})
\end{equation}
such that
\begin{equation}
\begin{aligned}
\mathbf{v} 
&\in \left[\, C_{\mathrm{weak}}\!\left([0,T]; L^2_{\mathbf{n}}(\Omega_\chi)\right)
          \cap L^2\!\left(0,T; W^{1,2}_{\mathbf{n}}(\Omega_\chi)\right) \,\right]^3, \\[0.3em]
\partial_t \mathbf{v} 
&\in \left[\, L^{4/3}\!\left(0,T; \left(W^{1,2}_{\mathbf{n}}(\Omega_\chi)\right)^*\right) \right]^3, \\[0.3em]
p 
&\in L^{4/3}\!\left(0,T; L^2_0(\Omega_\chi)\right), \\[0.3em]
\Bp 
&\in \left[\, C\!\left([0,T]; L^1(\Omega_\chi)\right) \right]^{3\times 3}
   \cap \left[\, L^2\!\left(Q_T\right) \right]^{3\times 3}, \\[0.3em]
\partial_t \Bp 
&\in \left[\, L^1\!\left(0,T; \left(W^{1,4}(\Omega_\chi)\right)^*\right) \right]^{3\times 3}, \\[0.3em]
\hat{\mathbf{u}} 
& \in \left[\, C\!\left(0,T; W^{1,2}(\Omega_\chi)\right) \right]^3
\end{aligned}
\end{equation}
and for all
\begin{equation}
(\boldsymbol{\phi},\psi,\mathbb{A},\hat{\mathbf{w}})\in [W^{1,2}(\Omega_\chi)]^3 \times L^2(\Omega_\chi) \times [W^{1,4}(\Omega_\chi)]^{3\times3} \times [W^{1,2}(\Omega_\chi)]^3
\end{equation}
the variational equations below are satisfied, assuming the ALE mapping remains orientation preserving, i.e., $\det \hat{\mathbb F} > 0$ a.e. in $\Omega_\chi$:
\begin{equation}
\begin{split}
\int_{\Omega_\chi} r\hat{J}\text{tr}\left((\nabla_{\hat{x}}\mathbf{v})\hat{\mathbb{F}}^{-1}\right)\psi = 0 &, \\
\Re\int_{\Omega_\chi}  r\hat{J}\left[\frac{\partial \mathbf{v}}{\partial t} + (\nabla_{\hat{x}}\mathbf{v})\left(\hat{\mathbb{F}}^{-1}\left(\mathbf{v} - \frac{\partial \hat{\mathbf{u}}}{\partial t}\right) \right) \right]\cdot\boldsymbol{\phi} + \int_{\Omega_\chi} r\hat{J}\hat{\mathbb{T}}\hat{\mathbb{F}}^{-T}\cdot\nabla_{\hat{x}}\boldsymbol{\phi} +\frac{1}{\Ca} \int_{\Gamma_\chi^\text{free}} r\|\hat{\mathbb{F}}^{-T}\mathbf{n}\|\left[
\mathbb{I}
-
\frac{\hat{\mathbb{F}}^{-T}\mathbf{n}}{\left|\hat{\mathbb{F}}^{-T}\mathbf{n}\right|}
\otimes
\frac{\hat{\mathbb{F}}^{-T}\mathbf{n}}{\left|\hat{\mathbb{F}}^{-T}\mathbf{n}\right|}
\right](\nabla_{\hat{x}}\boldsymbol{\phi})\hat{\mathbb{F}}^{-1}  &\\+ \frac{1}{\St} \int_{\Omega_\chi} r\hat{J} \mathbf{e}_z \cdot \boldsymbol{\phi} = 0 &, \\
\hat{\mathbb{T}} = -p\mathbb{I} + 2\frac{\mu_s}{\mu_s+\mu_p}\mathbb{D}_{\hat{x}} + \frac{a}{We}\frac{\mu_p}{\mu_s+\mu_p}(\Bp-\mathbb{I}) &, \\
\int_{\Omega_\chi} r\hat{J} \frac{\delta \Bp}{\delta t}\Big|_{\hat{x}} \cdot \mathbb{A} + \hat{J}\left[\frac{1-\alpha}{\We}(\Bp-\mathbb{I})+ \frac{\alpha}{\We}(\Bp^2-\Bp)\right]\cdot\mathbb{A} =0 &, \\
\int_{\Omega_\chi} r \, \partial_z \hat{u}_z \, \partial_z \hat{w}_z 
- \int_{\partial\Omega_\chi}r \, (\partial_z \hat{u}_z) \, \hat{w}_z 
- \int_{\partial\Omega_\chi} r \, \left[ \frac{\partial\hat{u}_z}{\partial t}  + (v_r \, \partial_r \hat{u}_z - v_z) \right]
\left( \partial_z \hat{w}_z - \frac{\beta_{st}}{h_\text{min}} \hat{w}_z \right)  = 0&,
\end{split}
\label{eq:dimless_weakForm_rodClimbing}
\end{equation}
where $\mathbf{n}$ is unit outward normal vector of the initial state, $\Re := \rho \omega R^2 / \mu$ is the Reynolds number, $\We := \tau \omega$ is the Weissenberg number, $\St := \mu \omega / (\rho g R)$ is the Stokes number, and $\Ca := \mu \omega R / \gamma$ is the capillary number. Regarding the physical constants, $R$ denotes the rod radius, $\omega$ the rotational frequency of the rod, $\gamma$ the surface tension coefficient, $g$ the gravitational acceleration, and $\mu := \mu_s + \mu_p$ the total viscosity, given as the sum of the solvent and polymer contributions. 

\section{Numerical implementation}\label{sec:numerics}

For the discretization of the governing equations, we employ the finite element method (FEM) and implement the formulation using the Firedrake library~\cite{FiredrakeUserManual}, which provides state-of-the-art infrastructure for automated FEM. The FEM is a natural choice for this problem: since we operate in a low-Reynolds-number regime, the Navier--Stokes equations behave effectively as a parabolic system, and we do not encounter the high-Weissenberg-number problem even without additional stabilization. Moreover, the FEM offers several advantages that are exploited in what follows.

We use higher-order elements because they provide excellent accuracy-to-cost performance \cite{parker2022high}. Their benefits are evident, for example, from the classical lid-driven cavity benchmark~\cite{Farrell2019} on Moffatt eddies, where higher-order elements have been shown to be particularly effective in capturing closed recirculation patterns \cite{Ainsworth2023}. Specifically, we employ the Scott--Vogelius pair of polynomial degree $p = 4$ ($[CG_p]^3$ for velocity and $DG_{p-1}$ for pressure) for the Navier--Stokes subsystem. This pair is provably stable on arbitrary 2D meshes and yields a pressure-robust method. Due to the axi-symmetric cylindrical weighting by the $r$-factor, and mainly due to the non-linear geometrical weighting by the a priori unknown deformation in the ALE method, pressure-robustness is not strictly preserved; nevertheless, the pair remained stable in all simulations we conducted. For the extra-stress tensor, we use $[CG_{p-1}]^{3 \times 3}_{\text{sym}}$, which matches the order of the pressure and the velocity gradient. For the mesh-displacement field, we use $CG_p$, since it must be of the same order as the velocity due to their direct coupling through the kinematic boundary condition.

When $p > 1$, the boundary geometry should also be represented in an isoparametric manner by polynomials of at least the same degree. Modern meshing tools such as Netgen, which is seamlessly integrated with Firedrake via ngsPETSc~\cite{Umberto2024}, now support this reliably. Choosing boundary polynomials of degree $p+1$ may in principle allow for superconvergence phenomena, although such behavior is not expected in our configuration. Importantly, our approach ensures that the free surface is represented at an arbitrary polynomial order: because we employ a virtual-ALE procedure and compute on a reference mesh, the physical mesh vertices remain fixed, and the boundary representation never degenerates to piecewise-linear, as would otherwise occur.

For time discretization, we adopt a three-step $\theta$-scheme of Glowinski~\cite{GLOWINSKI20033}, which is second-order accurate. In free-surface and moving-domain simulations, the preservation of fluid volume is critical. Our scheme does not guarantee exact volume conservation per se; instead, it relies on the accuracy of the incompressibility constraint ($\div \mathbf{v} = 0$) and the numerical preservation of the Jacobian determinant ($\int_{\Omega_\chi} \hat{J}(t) = \mathrm{const}$). In our tests, the scheme maintained excellent volume accuracy, whereas implicit Euler often did not for larger deformations. This choice also avoids unnecessary loss of temporal accuracy relative to our high-order spatial discretization.

Finally, we solve the fully coupled system in a monolithic fashion. This is particularly effective here: even with modest numbers of degrees of freedom (e.g., $130\,\mathrm{k}$), the high-order elements yield a highly accurate approximate solution since the actual solution to which we converge seems to be quite smooth. Moreover, monolithic coupling avoids artificially making the high-Weissenberg-number problem more severe, which may explain why no stabilization is required. The monolithic structure also enables the ``lagged-Jacobian'' strategy, allowing several Newton iterations to be performed almost for free by reusing a stored LU factorization. With sufficiently small time steps, the non-linear iteration count remains low. In our simulations up to final time $T = 5\,\mathrm{s}$, we use $\Delta t = 0.01\,\mathrm{s}$; on average, five Newton iterations per step suffice, and we lag the Jacobian for 50 iterations, greatly reducing computational cost. For example, when steady state is reached at $T_\text{st} = 2$ s, the linear system only needs to be assembled and factorized approximately 20 times over this time interval.

Overall, the method achieves high accuracy; the implementation runs in parallel with short wall-clock times (10 minutes for the high-accuracy $p=4$ setting with $130\,\mathrm{k}$ DoF, and 4 minutes for the $p=2$ setting with $23\,\mathrm{k}$ DoF, both on an average laptop) and is efficient in all essential aspects. The main limitation arises from large non-smooth deformations of the free surface under the virtual-ALE mapping, which would, in more extreme settings, require more intricate mesh motion or even remeshing during time-stepping. Introducing such operations would require particular care in order to preserve the expected convergence order of the solution. For this reason, we restrict our attention to regimes in which the free-surface deformation remains sufficiently smooth and can be handled robustly within the present ALE framework. Within this range of parameters, no additional mesh motion or remeshing is required. We note that the method is not intended for regimes involving topological changes or self-contact of the free surface, such as the breathing-instability regime reported in \cite{FIGUEIREDO201698}.

On the other hand, a major advantage over Eulerian multiphase formulations is that no specialized numerical treatment of the free interface is required, as is the case, for example, in Figueiredo et al.~\cite{FIGUEIREDO201698}. The interface is handled naturally, without reconstruction procedures, interface-capturing enhancements, or CFL-type restrictions on interface motion \cite{GARCIAVILLALBA2025105285}. Moreover, owing to the fully implicit time discretization and the monolithic solution strategy, the temporal evolution is resolved with high accuracy and without introducing additional splitting or coupling errors. As a result, the numerical time evolution closely follows the expected convergence toward the steady state, and the computed climbing-profile shape of the free surface is captured with high precision. %

\section{Results}\label{sec:results}
In this section, we simulate the rod climbing experiment in the regime $\Re \ll 1$ and $\We < 1$; however, different regimes may be accessible with our code \cite{cach_rod_climbing2026}. We follow the configuration studied in \cite{FIGUEIREDO201698}, which in turn connects to a series of real experiments conducted by Beavers and Joseph in 1975~\cite{Beavers_Joseph_1975} and to numerical results reported by Luo in 1999~\cite{LUO1999393}. We compare our computations with these earlier works and extend the analysis by presenting new results for a broader class of viscoelastic constitutive models. As part of the validation, we also document the convergence of the free-surface shape. Specifically, we examine mesh refinement for the widely used choice $p = 2$, as well as $p$-refinement on a moderately fine mesh (M1, see the Oldroyd-B convergence study below). As we will show, the best surface representation obtained with our high-order FEM is comparable to that obtained on much finer meshes with classical low-order Taylor--Hood elements \cite{Taylor1973}, but at a significantly reduced computational cost.

Our comparison with previous studies focuses on the climbing-profile shapes obtained for the Oldroyd-B and Johnson--Segalman models, the latter having been used to fit experimental data for real polymer solutions in \cite{LUO1999393}.
However, its ability to reproduce the rod climbing experiment is limited for two reasons. First, the steady climbing height depends strongly on the JS slip parameter, which is itself highly temperature-dependent; even small temperature variations cause noticeable changes in the predicted height. Although this issue is purely practical and not present in our isothermal computations, it can be addressed by careful parameter fitting. A second, more fundamental limitation concerns the shape of the steady free surface. The engineering JS model appears to produce a similar shape for small changes in the slip parameter, but it greatly affects climbing height. This suggests a structural limitation of the model itself to fit experimental data; see again the work \cite{LUO1999393}. This points to the need for either alternative models or thermodynamically consistent generalizations of the commonly used ones. In the Rajagopal--Srinivasa framework, such generalizations often arise naturally; for example, through different admissible choices of the rate of dissipation.

\subsection{Configuration}\label{sec:configuration}
The experimental and numerical configurations reported in the literature differ in several aspects. In this work, we adopt the configuration as in \cite{FIGUEIREDO201698}, although the distance between the container bottom and the lower end of the rod is not specified therein. The choice of boundary conditions is likewise non-trivial and is only briefly discussed in the available numerical studies. For instance, the work \cite{LUO1999393} imposes a non-penetration and free-slip condition on the outer radial boundary, whereas \cite{FIGUEIREDO201698} employ a full no-slip condition. Although this boundary is located away from the primary region of interest, such differences may still have a slight impact on the resulting flow. A more significant modeling ambiguity arises at the surface of the rotating rod, where different assumptions are made. While imposing free-slip in the vertical direction, the no-slip condition is enforced in the rotating direction. We are not sure if such ``anisotropy'' is physically plausible. In experimental studies, further variability is introduced by differences in rod materials and surface coatings, which are known to affect the observed climbing height. Regarding the boundary condition on the rod, we note that the commonly adopted non-penetration condition, typically enforced as a Dirichlet constraint $v_r = 0$, additionally precludes fluid detachment from the rod surface. In the low-Reynolds-number regime considered here, this assumption appears physically reasonable.

\begin{figure}[h!]
\centering
\begin{tikzpicture}[scale=0.6]

\def\R{1}        %
\def\Htop{12}    %
\def\Hwall{10}   %
\def\Rmax{12}    %

\draw[dashed, thick,->] (0,0) -- (0,14*\R) node[right] {$z$};
\draw[thick,->] (0,0) -- (14*\R,0) node[above] {$r$};
\node at (-0.35*\R,-0.15*\R) {$\mathbf{0}$};

\draw[thick, ->,domain=200:520, samples=80] plot ({1.3*\R*cos(\x)}, {\Htop*\R + 0.9*\R*sin(\x)});
\node at (1.4*\R,\Htop*\R+\R) {$\omega$};

\draw[thick] (\R,\R) -- (\R,\Htop*\R);
\draw[thick] (-\R,\R) -- (-\R,\Htop*\R);
\node[rotate=90, anchor=center]
  at (1.5*\R,0.5*\R+0.4*\Htop*\R) {rotating rod mantle};

\draw[thick] (0,\Htop*\R) ellipse (1 and 0.5);

\draw[very thick] (\Rmax,0) -- (\Rmax,\Hwall*\R);
\node[right] at (\Rmax,0.5*\Hwall*\R) {wall};

\draw[very thick] (0,0) -- (\Rmax*\R,0);
\node[below] at (0.5*\Rmax*\R,0) {wall};

\node[left] at (-\R,\R) {$z=2R$};
\node[left] at (-\R,\Htop*\R) {$z=\infty$};

\draw[thick,<->] (0,4*\R) -- (\R,4*\R);

\draw[thick] (-\R,\R) arc (180:360:1 and 0.5);
\draw[dashed] (-\R,\R) arc (180:0:1 and 0.5);

\node[below] at (0.5*\R,4*\R) {$R$};

\draw[thick,<->] (0,\R) -- (\Rmax,\R);
\node[above] at (0.5*\Rmax,\R) {$R_c = 24R$};

\draw[dashed, thick] (\R,0.9*\Hwall) -- (\Rmax,0.9*\Hwall);
\node[above] at (0.5*\Rmax,0.9*\Hwall) {initial free surface};

\draw[thick,<->] (\Rmax-\R,0.9*\Hwall) -- (\Rmax-\R,0);
\node[left] at (\Rmax - \R,0.5*0.9*\Hwall) {$H_0 = 12R$};

\draw[thick,<->] (\R,0.9*\Hwall) -- (\R,0.9*\Htop);
\node[left] at (\R,0.5*0.9*\Hwall+0.5*0.9*\Htop) {$H$};

\pgfmathsetmacro{\zmatch}{0.9*\Hwall}
\pgfmathsetmacro{\zcenter}{(0.9*\Htop+\zmatch)/2}
\pgfmathsetmacro{\ampl}{(0.9*\Htop-\zmatch)/2}
\pgfmathsetmacro{\depth}{0.25*\R}

\draw[thick, domain=\R:3*\R, samples=100]
plot ({\x},
      {\zcenter + \ampl * cos(deg(pi*(\x-\R)/(2*\R)))});

\pgfmathsetmacro{\xm}{(3*\R+\Rmax)/2}
\pgfmathsetmacro{\a}{\depth/((\Rmax-\xm)^2)}

\draw[thick, domain=3*\R:\Rmax, samples=100]
plot ({\x},
      {\zmatch-\depth + \a*(\x-\xm)^2});

\end{tikzpicture}

\caption{Sketch of the axi-symmetric rod climbing configuration used in all simulations. The problem is formulated in the meridional plane of cylindrical coordinates, with a rotating rod, a deformable free surface, and no-slip container walls.}
\label{fig:rod_climbing_setup}

\end{figure}
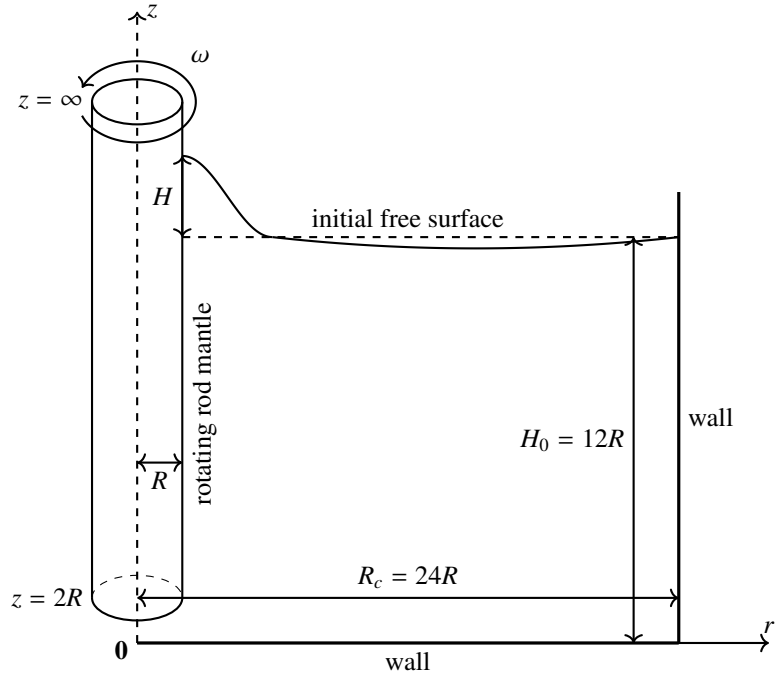

\Cref{fig:rod_climbing_setup} shows the rod-climbing configuration considered in this work. A rod of radius $R = 0.635$~cm and infinite height rotates at a constant angular velocity $\omega$ inside a cylindrical container of radius $R_c = 24R$, with its lower end located at height $z = 2R$ above the container bottom. The domain is described in cylindrical coordinates, and the governing equations are solved in the meridional plane under the axi-symmetric assumption. The container walls (bottom and outer cylinder) are subject to no-slip boundary conditions, while the free surface is initialized at a height $H_0 = 12R$ and evolves according to the kinematic condition. Non-penetration is enforced everywhere, and the fluid is assumed to be perfectly advected by the rod in the azimuthal direction. To allow for vertical climbing, free-slip is imposed in the axial direction along the rod mantle, while full no-slip is enforced at the rod bottom. The small region beneath the rod bottom is treated as a symmetry boundary, enforcing zero radial velocity, free-slip in the axial direction, and zero azimuthal velocity.

The fluid parameters specified in the simulations, see the weak form of Model~I (\ref{eq:dimless_weakForm_rodClimbing}), are the slip parameter $a$, the mobility parameter $\alpha$, and the dimensionless numbers: Reynolds ($\Re$), Weissenberg ($\We$), Stokes ($\St$), and capillary ($\Ca$). The total viscosity is fixed to $\mu = 14.6~\mathrm{Pa\,s}$, with the solvent-to-polymer viscosity ratio given by $\mu_s / \mu_p = 1/9$.

For the special cases of the Oldroyd--B and Johnson--Segalman models, the material parameters are set to $\rho = 890~\mathrm{kg\,m^{-3}}$, $\gamma = 0.0308~\mathrm{N\,m^{-1}}$, $g = 9.81~\mathrm{m\,s^{-2}}$, and a relaxation time $\tau = 0.0162/(1 - \mu_s / \mu_p)$. These parameters are subsequently expressed in terms of the dimensionless groups defined above for different values of the angular velocity $\omega$. In all simulations, a steady state is reached well within a final simulation time of $T = 5~\mathrm{s}$.

In the sections concerning the Giesekus model, additional dimensionless numbers are employed, which can be expressed in terms of the primary ones. These include the Elastic number $\El = \We / \Re$, the Bond number $B = \Ca / \St$, the elastic gravity number $\mathcal{G} = \We / \St$, and the polymer viscosity ratio $\beta_p = \mu_p / \mu$. Unless stated otherwise, we keep the same parameter values as in the other sections.

\subsection{Oldroyd-B model}
In this subsection, we validate our numerical implementation of the Oldroyd-B model ($a = 1$, $\alpha = 0$, common limit of both models (\ref{eq:modelA}) and (\ref{eq:modelB})) by comparing our results with those reported in \cite{FIGUEIREDO201698}, for which the same constitutive model and physical setup are considered. As shown previously, the Oldroyd-B model can be obtained within Rajagopal-Srinivasa's thermodynamic framework and is consistent with the second law of thermodynamics.

\begin{figure}[h!]
    \centering
    \begin{subfigure}[b]{0.495\textwidth}
        \centering
        \includegraphics[width=\textwidth]{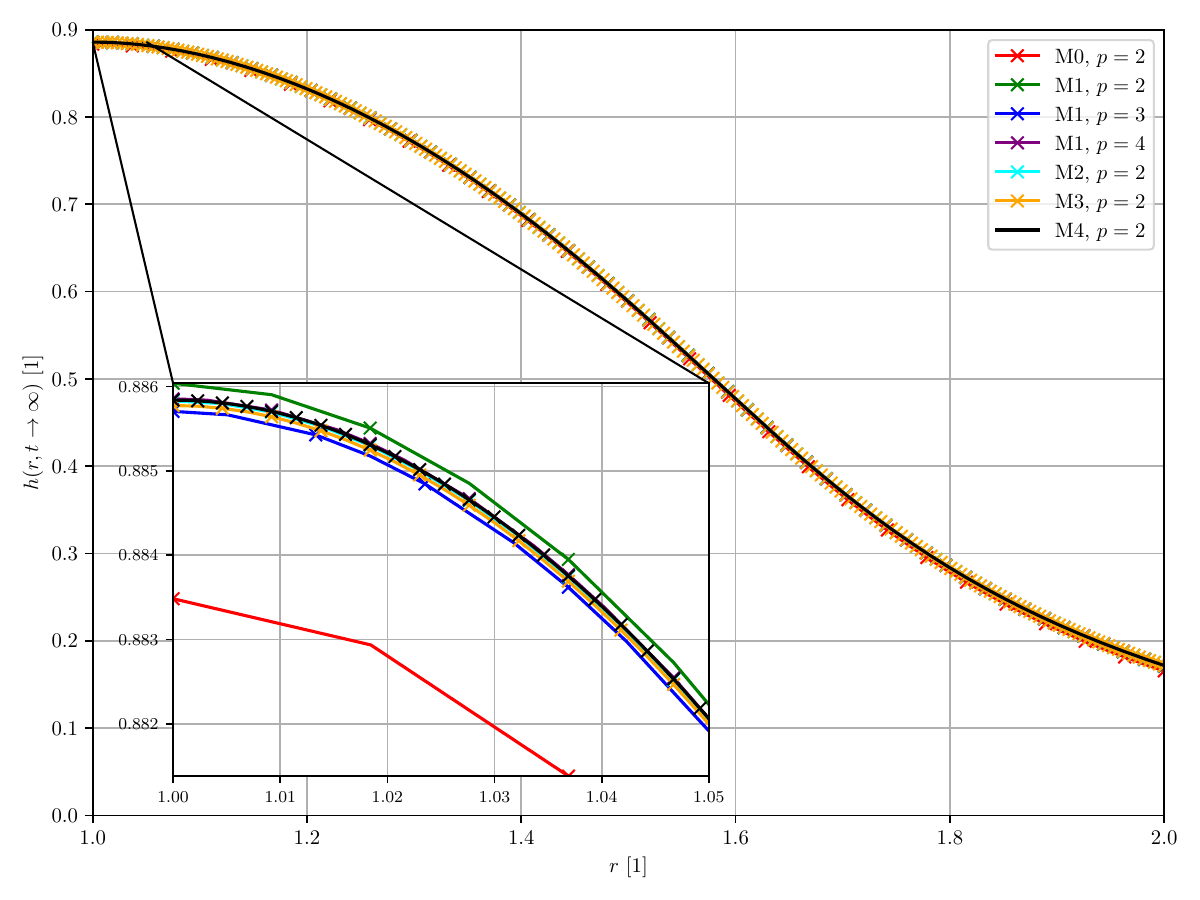}
        \caption{Comparison of mesh- and p-convergence: shapes. Every vertex DoF is depicted by marker.}
        \label{fig:oldroyd_b_convergence}
    \end{subfigure}
    \hfill
    \begin{subfigure}[b]{0.495\textwidth}
        \centering
        \includegraphics[width=\textwidth]{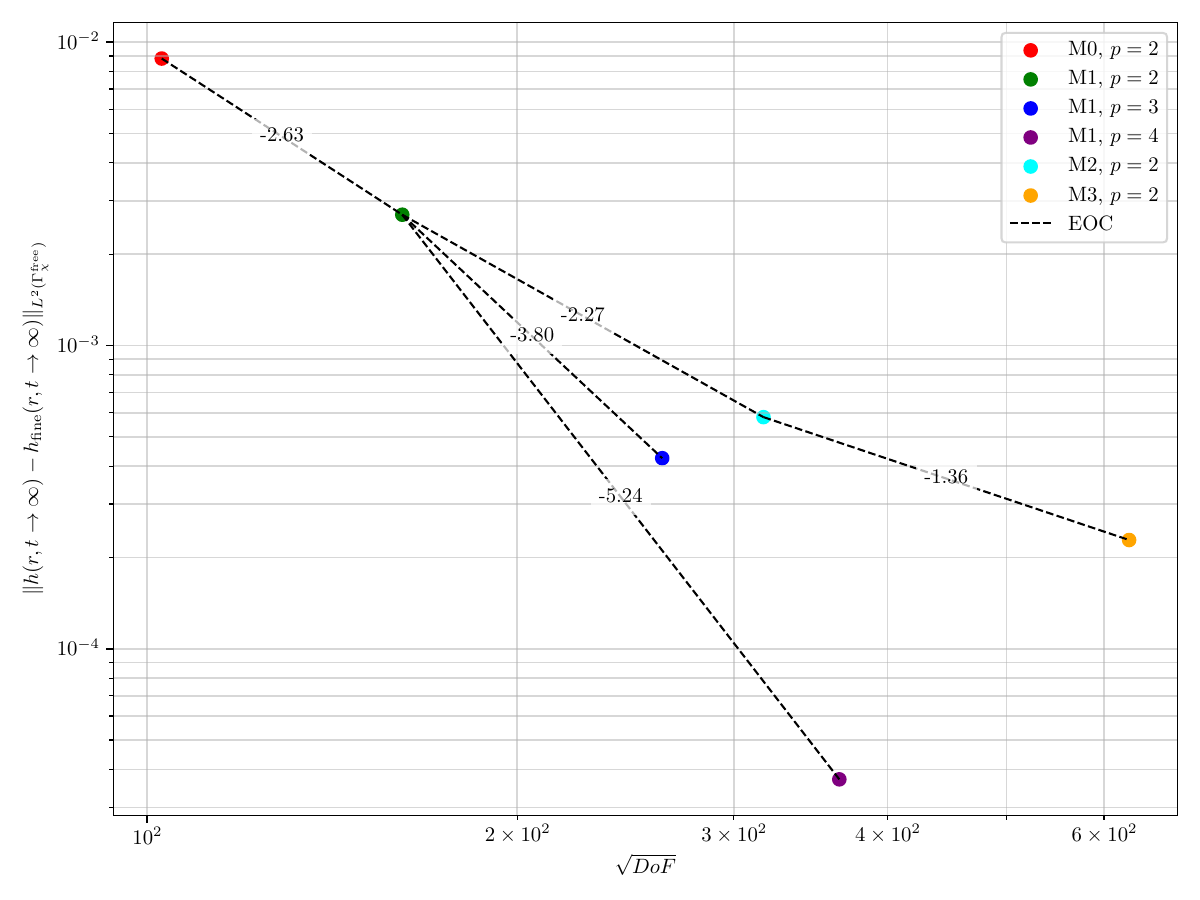}
        \caption{Comparison of mesh- and p-convergence: $L^2$ log-log errors w.r.t.~mesh M4 with $p=2$ (1.5M DoF) result on the free-surface boundary $\Gamma_\chi^\text{free}$. Estimated order of convergence (EOC) locally depicted.}
        \label{fig:oldroyd_b_convergence_formal}
    \end{subfigure}
    \caption{Spatial discretization convergence is examined for $\omega = 2.6$ rev/s. The meshes are labelled M0, \dots, M4 from coarsest to finest. Mesh M1 with $p=4$ (approximately $1.3\times 10^{5}$ DoF) produces results essentially identical in the region near the rod to those obtained on mesh M4 with $p=2$ (approximately $1.5\times 10^{6}$ DoF), demonstrating that higher-order discretizations achieve same accuracy at much lower computational cost. The convergence is not strictly monotone, with small oscillations around what appears to be the mesh-converged solution.}
    \label{fig:oldroyd_b_convergence_full}
\end{figure}

Before presenting quantitative comparisons, we assess the numerical robustness of our formulation through a mesh and polynomial refinement study. See \Cref{fig:oldroyd_b_convergence_full} for mesh ID notation, the associated polynomial degree $p$ of the SV pair, and the overall DoF count. This study verifies that the computed solutions are independent of the spatial discretization and that the observed flow features are not affected by numerical artifacts. The results demonstrate that the ALE formulation retains the expected convergence properties of standard FEM. Even on the second coarsest mesh, the solution already provides a good approximation of the free-surface profile. This is likely because the flow pattern is not very complicated and the free surface near the rod is meshed sufficiently. The flow pattern consists of a large secondary recirculation over the whole domain, with a smaller recirculation in the rod climbing region. The latter area was a priori refined with a $50\times$ focus, which explains why the coarse meshes already perform well. With increasing mesh resolution and a higher polynomial degree, the solution improves as one would wish: the free surface exhibits only a very small variations in point-wise climbing height, indicating mesh- and $p$-independent behavior.

\begin{figure}[h!]
    \centering
    \begin{subfigure}[b]{0.495\textwidth}
        \centering
        \includegraphics[width=\textwidth]{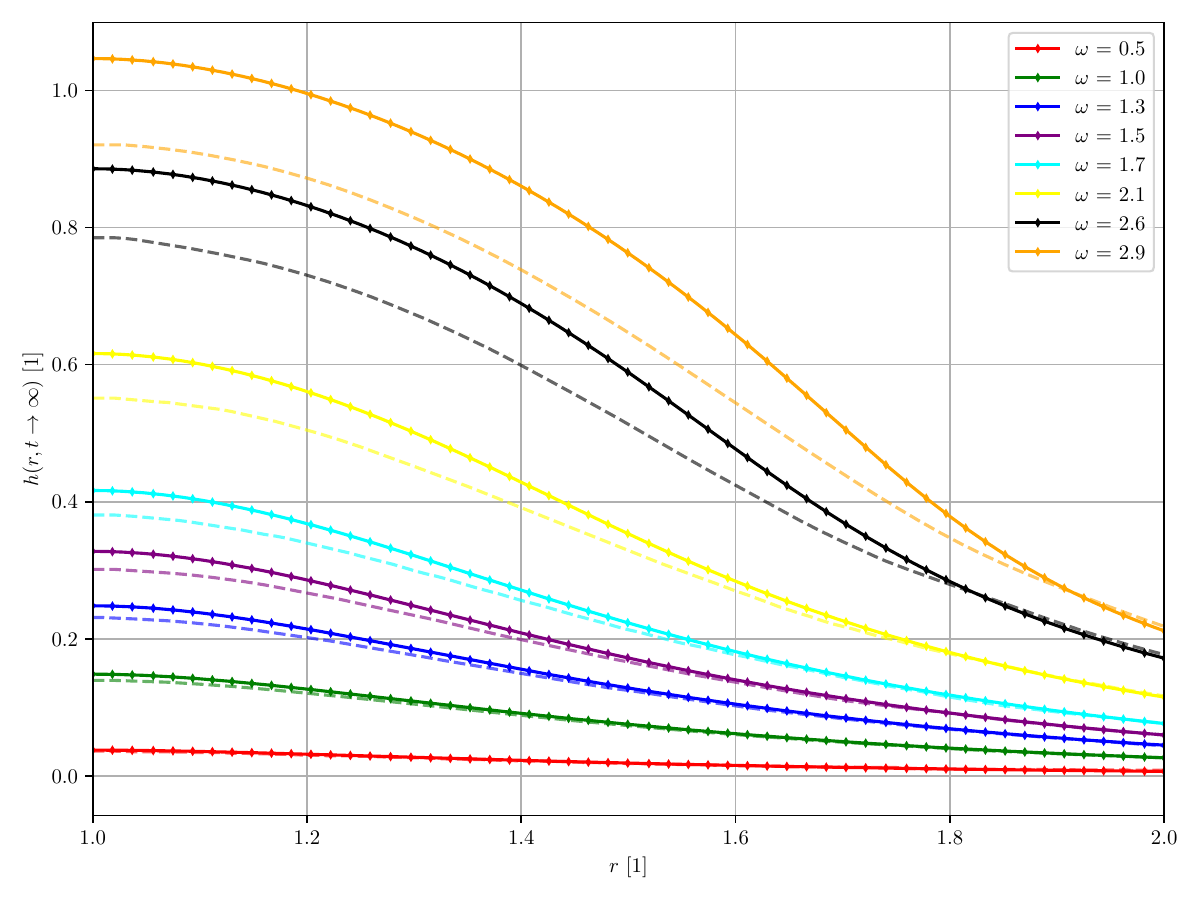}
        \caption{Full lines are computed in this study, with every 4th vertex DoF depicted with markers, and the piecewise quartic solution in between. Dashed line by \cite{FIGUEIREDO201698}, extracted by automeris.io \cite{WebPlotDigitizer}.}
        \label{fig:oldroyd_b_comparison}
    \end{subfigure}
    \hfill
    \begin{subfigure}[b]{0.495\textwidth}
        \centering
\begin{tabular}{c c c c c}
\toprule
$\omega$ (rev/s) & $\We$ & $\Re$ & $\St$ & $\Ca$ \\
\midrule
0.5 & 0.057 & 0.008 & 0.827 &  9.43 \\
1.0 & 0.115 & 0.015 & 1.655 & 18.85 \\
1.3 & 0.149 & 0.020 & 2.151 & 24.51 \\
1.5 & 0.172 & 0.023 & 2.482 & 28.28 \\
1.7 & 0.195 & 0.026 & 2.813 & 32.05 \\
2.1 & 0.240 & 0.032 & 3.475 & 39.59 \\
2.6 & 0.298 & 0.040 & 4.302 & 49.01 \\
2.9 & 0.332 & 0.045 & 4.798 & 54.67 \\
\bottomrule
\end{tabular}

        \caption{Dimensionless numbers of the rotation rates used and corresponding dimensionless numbers in the simulations. These apply for all models, as $a$ and $\alpha$ are independent parameters.}
        \label{tab:oldroyd_b_table}
    \end{subfigure}

    \caption{Comparison of the Oldroyd-B results obtained in the present work with the reference results \cite{FIGUEIREDO201698}. The observed differences may partly be attributed to spatial under-resolution in the reference simulations, as suggested by the coarse structure of the reported climbing vortex.}
    \label{fig:climbing_oldroyd}
\end{figure}

\begin{figure}[h!]
    \centering
    \includegraphics[width=0.995\linewidth]{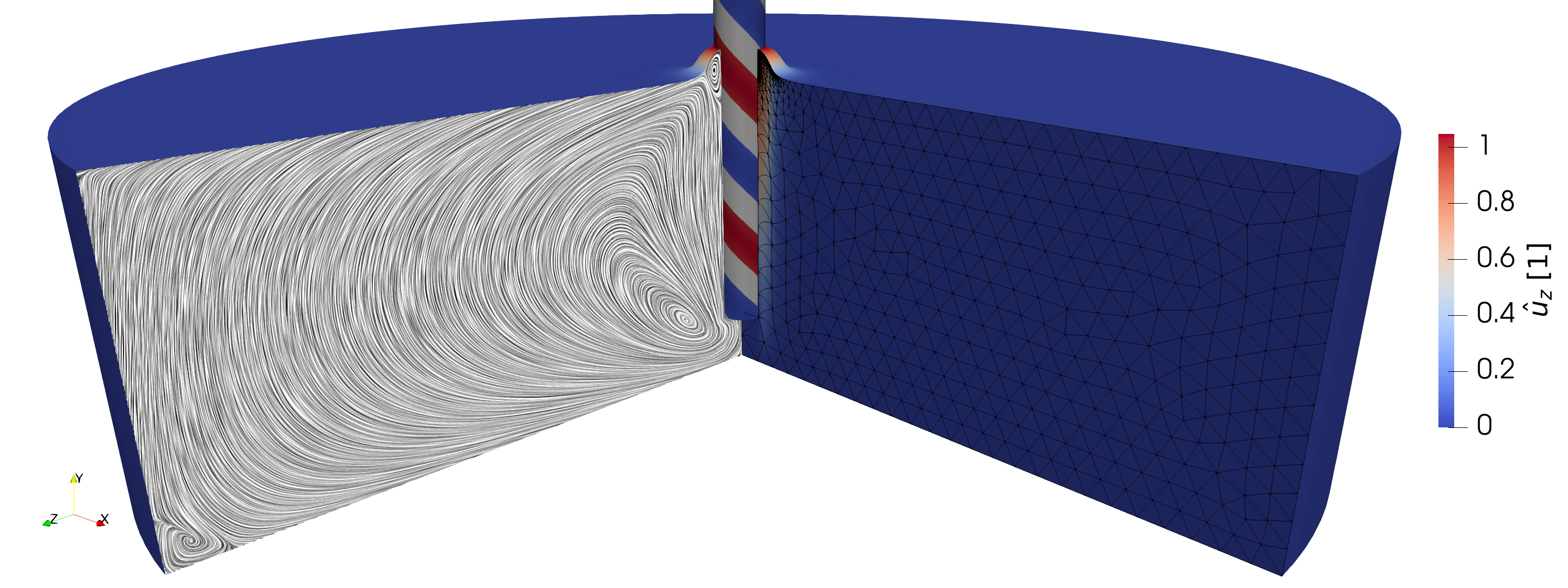}
    \caption{Visualization of the resulting steady state for $\omega = 2.9$~rev/s in ParaView. The domain is colored and warped by the computed mesh displacement $\hat{\mathbf{u}} = (0,0,\hat{u}_z)$ and shown in a three-dimensional perspective obtained by a $270^\circ$ rotation about the rotation axis. The right face cut shows the computational mesh (M1), while the left face cut displays a line integral convolution (LIC) representation of the velocity field $\mathbf{v} = (v_r,0,v_z)$ in the meridional plane, showing the secondary flow. The rotating rod, shown for reference, is not a part of the simulation; its motion is prescribed via boundary conditions.}
    \label{fig:3d_revolution}
\end{figure}

With the mesh-converged solutions established, we compare the predicted free-surface deformation with the reference results. We use the mesh M1 with $p=4$, as introduced in the numerics section. Overall, we systematically observe a noticeably higher climbing height, see \Cref{fig:climbing_oldroyd}, with the difference becoming more pronounced at higher rotation speeds. For the lowest rotation speed, the results coincide with the reference, suggesting that the difference occurs for large Weissenberg numbers. This discrepancy does not necessarily indicate a numerical issue, but may instead arise from differences in the configuration setup, specifically due to the boundary conditions on the rod and on the right wall, or the rod depth (distance from the bottom), as discussed in \Cref{sec:configuration}. In \Cref{fig:3d_revolution}, we depict the resulting steady state for $\omega = 2.9$ rev/s in ParaView by rotating the axi-symmetric solution warped by the mesh deformation $\hat{\mathbf{u}}$ about the symmetry axis, yielding a representative three-dimensional visualization of the rod climbing fluid shape. The line integral convolution visualization of the velocity field in the meridional plane reveals a coherent vortex structure in the climbing region.

In terms of the maximal climbing height, defined as $H := h(r=a,\, t \to \infty)$, the qualitative behavior agrees with the reference results \cite{FIGUEIREDO201698}; however, our simulations indicate a more rapid growth of $H$.
This observation is consistent with the findings reported in the reference, where the mesh refinement study shows that the predicted climbing height continues to increase with increasing spatial resolution and is not yet fully converged at the resolution employed. Extrapolating these trends suggests that the two results may not differ as significantly as initially observed.

It is worth noting that convergence of boundary values in finite-element approximations is generally slower than convergence in the bulk (heuristically of order $p-\tfrac{1}{2}$ in the $H^1$ norm) and depends critically on the regularity of the associated adjoint problem. For viscoelastic flows governed by the Oldroyd-B model, such regularity cannot be expected in general. Consequently, first-order spatial discretizations may exhibit very slow or negligible convergence of boundary quantities, underscoring the importance of higher-order spatial discretizations. The same issue arises in computing the boundary traction of solutions to the Navier--Stokes equations with high-accuracy, as reported in recent work~\cite{cach2025}.

\subsection{Johnson--Segalman Models I and II}
In this section, we consider two closely related formulations of the Johnson--Segalman model (here with $\alpha = 0$): the classical engineering form (\ref{eq:modelB}) and a thermodynamically consistent variant (\ref{eq:modelA}) compatible with the Rajagopal--Srinivasa framework. Both formulations generalize an Oldroyd-B type description by introducing a ``polymer chains slip parameter'' $a$ through a modified objective derivative acting on the conformation tensor. However, the two models differ in the manner in which elastic stresses are coupled to the kinematics.

In the engineering formulation, the polymeric stress contribution retains the form $G(\mathbb{B}-\mathbb{I})$ independently of the value of $a$, while the slip parameter enters only through the evolution equation for the conformation tensor. In contrast, the thermodynamically consistent formulation scales the elastic stress contribution by the factor $a$, such that both the stored elastic energy and the associated stresses vanish continuously as $a \to 0$. As a result, the latter model admits a well-defined Newtonian limit, while the former does not enforce such a limit at the level of the stress response.

\begin{figure}[h!]
    \centering
    \begin{subfigure}[b]{0.495\textwidth}
        \centering
        \includegraphics[width=\textwidth]{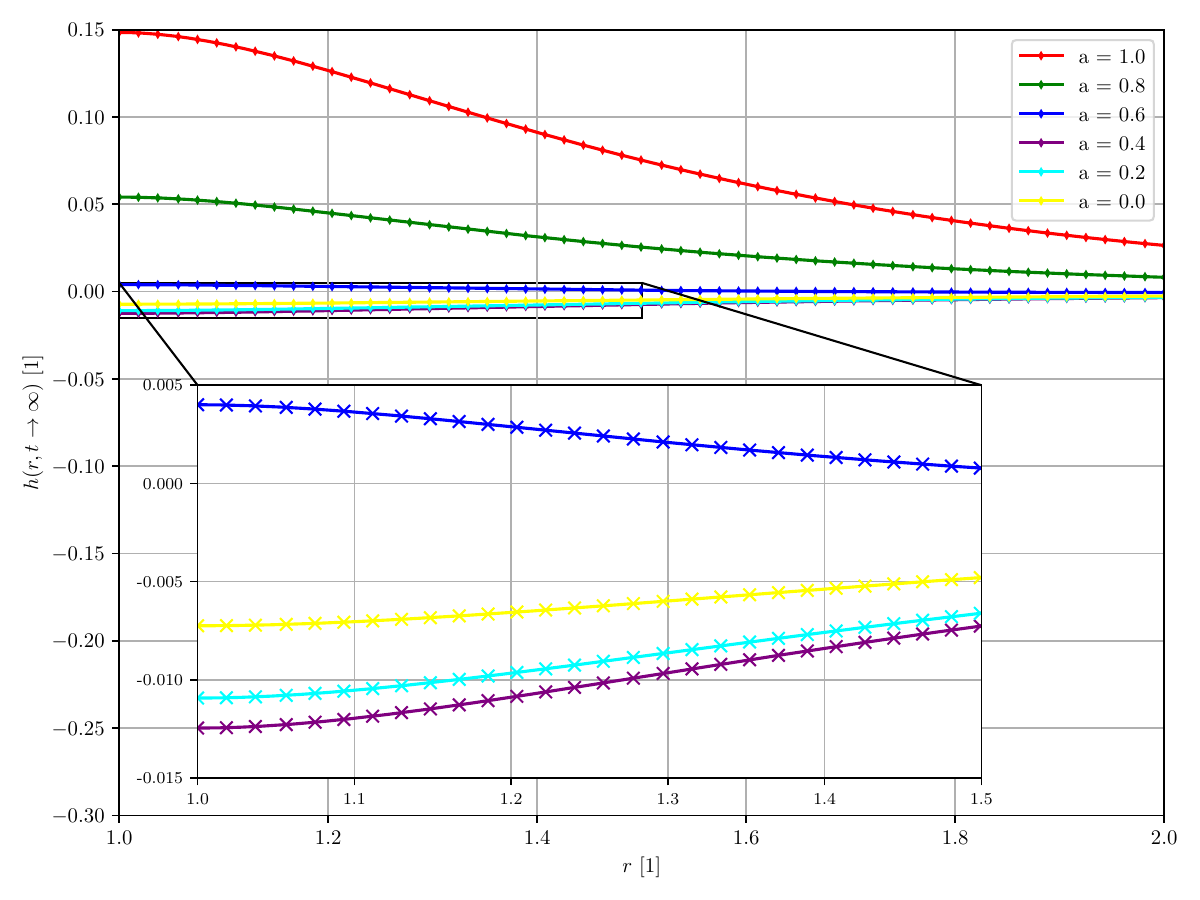}
        \caption{Model~I, every 4th vertex DoF depicted with markers, and the piecewise quartic solution in between.}
        \label{fig:JS_A}
    \end{subfigure}
    \hfill
        \begin{subfigure}[b]{0.495\textwidth}
        \centering
        \includegraphics[width=\textwidth]{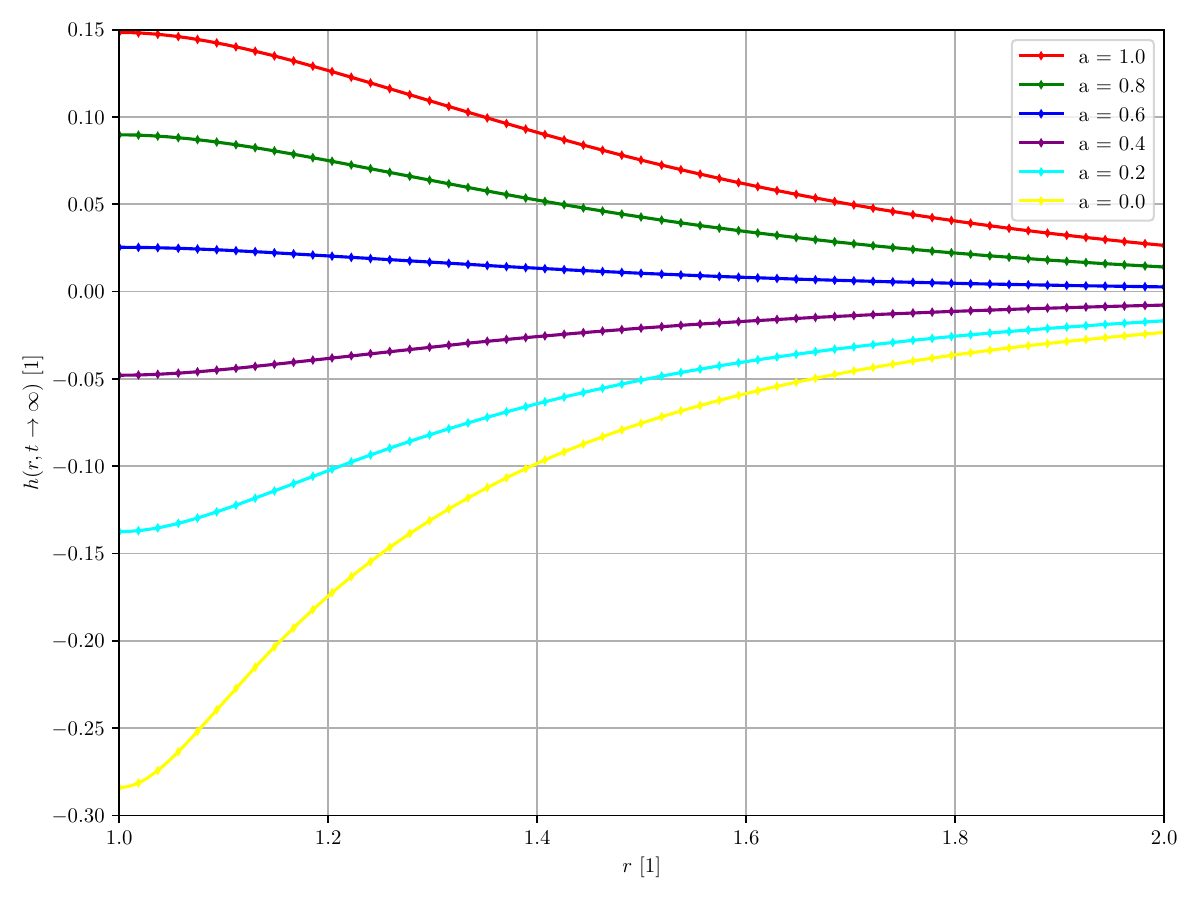}
        \caption{Model~II, every 4th vertex DoF depicted with markers, and the piecewise quartic solution in between.}
        \label{fig:JS_B}
    \end{subfigure}
    \caption{Comparison of results obtained in the present work for JS Models I and II at $\omega = 1.0$~rev/s. The Model~I solution approaches the Newtonian limit as $a \to 0$ in a non-monotonic manner, with a regime of $a$ in which rod descending is slightly stronger than in the purely Newtonian case. In contrast, rod descending is greatly intensified in Model~II.}
    \label{fig:JS_AB_comparison}
\end{figure}

To assess the physical implications of this structural difference, we compare the predictions of both formulations in the same rod climbing setup described in \Cref{sec:configuration}. In the limiting case $a=1$, both models reduce to the Oldroyd--B model and yield the same rod climbing behavior. As the slip parameter $a$ is reduced, the predictions of the two models diverge significantly. In the thermodynamically consistent formulation (Model~I), the elastic stresses weaken proportionally with $a$, and the flow transitions smoothly toward a Newtonian response characterized by a slight rod descending behavior governed by viscous and inertial effects. In contrast, the engineering JS (Model~II) predicts a regime of pronounced rod descending for intermediate values of $a$, with steep free-surface gradients developing near the rotating rod. Even though Model~I also contains a regime in which the rod descending is slightly enhanced, it is small compared to the surface depression of Model~II, which is even larger than the magnitude of the climb. See \Cref{fig:JS_AB_comparison} for the surface shape plots at $\omega = 1.0$ rev/s (remaining parameters in \Cref{tab:oldroyd_b_table}) over $a\in[0,1]$.

From a phenomenological perspective, such enhanced rod descending behavior is not supported by experimental observations of viscoelastic fluids. Rod climbing and rod descending experiments are well described by combinations of low-shear normal stress coefficients. While weak rod descent may occur when inertial effects dominate, to the best of our knowledge, there is no evidence that viscoelastic fluids exhibit strongly amplified rod descending responses of elastic origin. The absence of experimental support for this regime, together with the lack of a Newtonian limit in the engineering formulation, suggests that the predicted behavior arises from the constitutive structure of the model rather than from physically admissible material responses.

\begin{figure}[h!]
    \centering
    \begin{subfigure}[b]{0.495\textwidth}
        \centering
        \includegraphics[width=\textwidth]{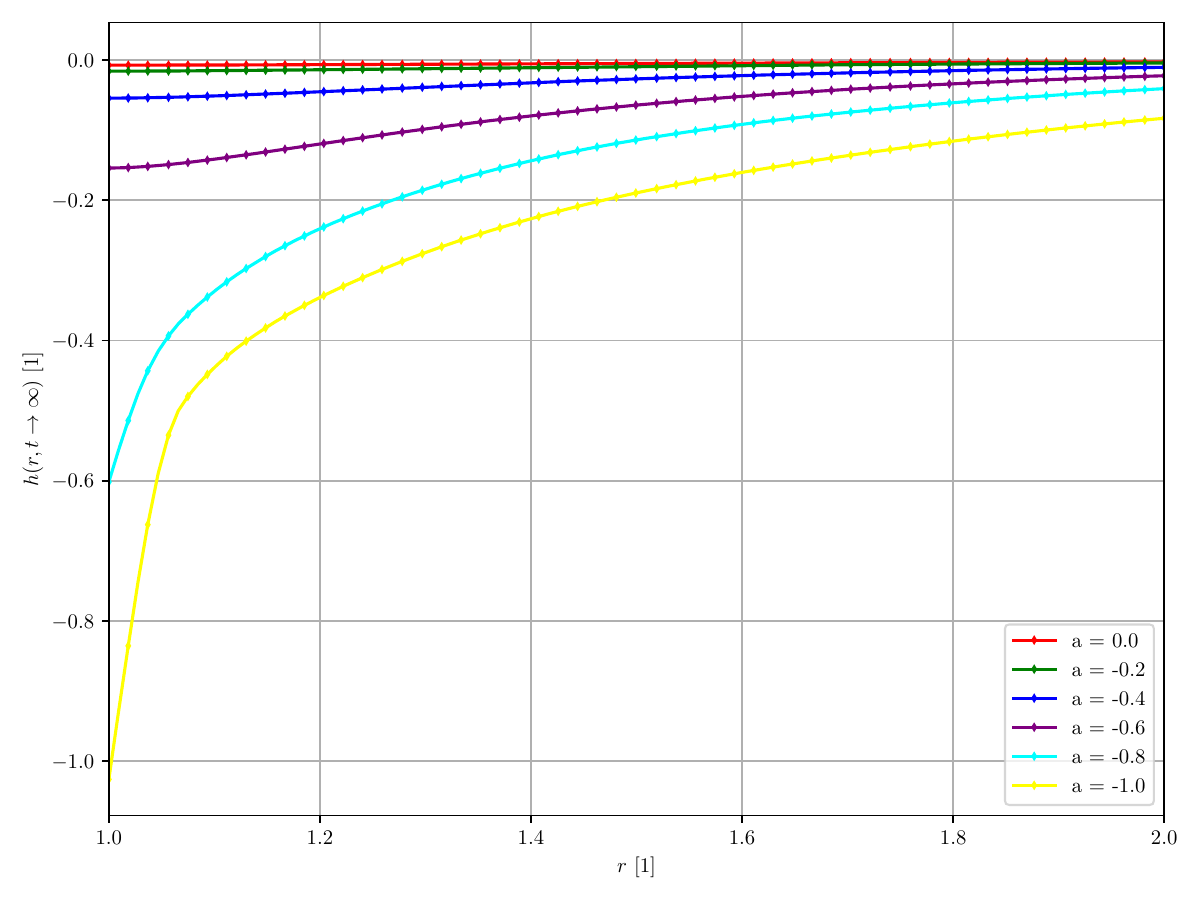}
        \caption{Model~I, every 2nd vertex DoF depicted with markers, and the piecewise quadratic solution in between.}
    \end{subfigure}
    \hfill
        \begin{subfigure}[b]{0.495\textwidth}
        \centering
        \includegraphics[width=\textwidth]{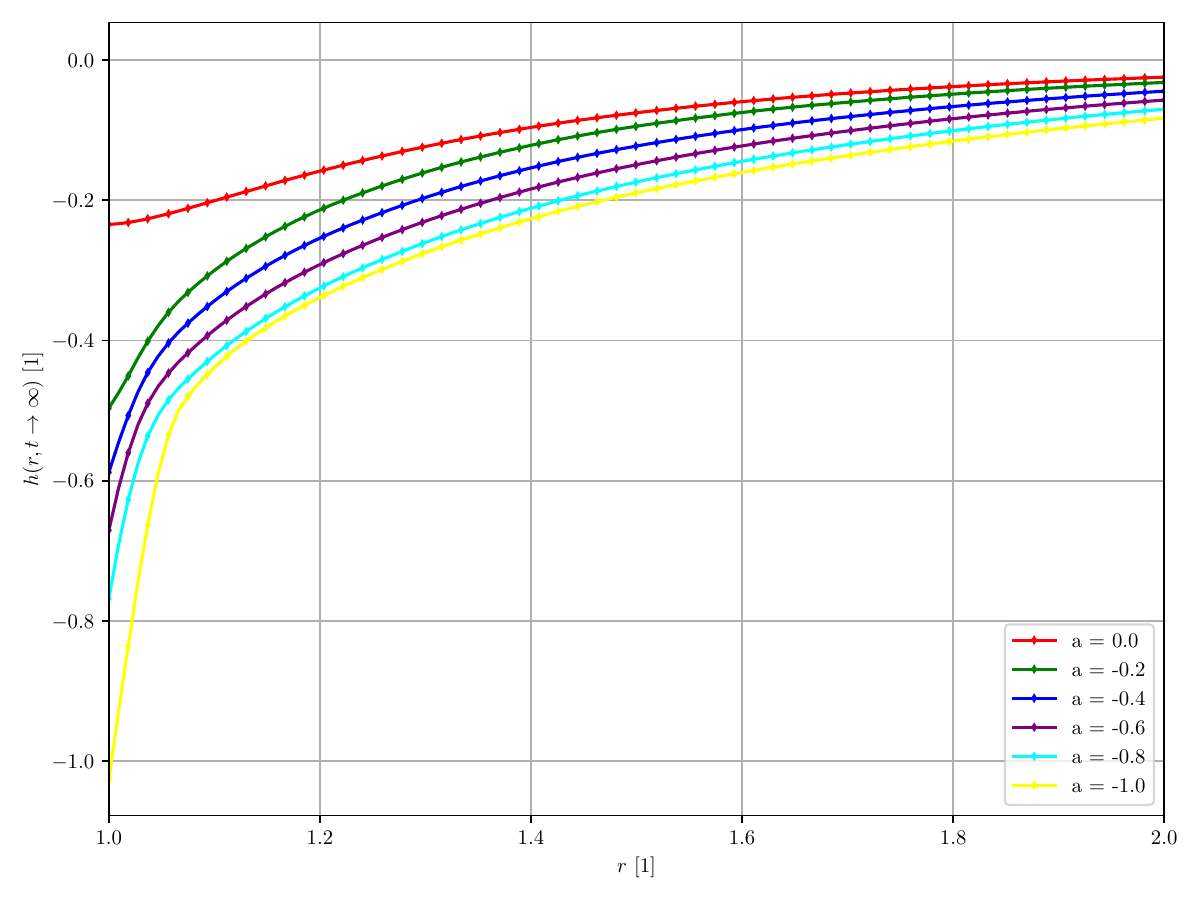}
        \caption{Model~II, every 2nd vertex DoF depicted with markers, and the piecewise quadratic solution in between.}
    \end{subfigure}
    \caption{Comparison of the results obtained in the present work for JS Models~I and~II at $\omega = 1.0$~rev/s. The Model~I results diverge from the Navier--Stokes solution for $a<0$, leading to pronounced rod descending, similarly to Model~II. At $a=-1$, both models coincide, corresponding to the commonly known Oldroyd--A model.}
    \label{fig:Oldryod_A}
\end{figure}

The mathematically admissible regime of the Gordon--Schowalter derivative (\ref{eq:GSder}) for $a \in [-1,0]$ remains an open question in terms of physical justification \cite{HINCH2021104668}. In this work, we assess it from the perspective of macroscopic behavior. We confirm that, in this regime, the model does not produce rod climbing, but instead leads to significant rod descending\footnote{Rod descending induces mesh compression in the ALE framework, in contrast to rod climbing, which predominantly stretches the mesh. Compression is particularly detrimental, since even moderate displacements may cause element inversion once a triangle is compressed to the scale of $h_{min}$, if not adequately accommodated by the bulk mesh motion. Under stretching, elements can typically tolerate larger deformations, with the primary risk being the development of near-singular angles. In addition, in our setting rod descending leads to free surfaces with much steeper slopes, which are especially challenging for our ALE implementation. To mitigate these effects, we enhance the bulk mesh motion by introducing a spatially variable coefficient in the Laplace equation governing the mesh displacement, chosen based on the local mesh size. Nevertheless, rod descending remains less robustly resolved in the present implementation; in regimes exhibiting strong descending, mesh compression may lead to element inversion, in which case the solver typically fails at the corresponding time step.}; see results in \Cref{fig:Oldryod_A}. Specifically, Model~I diverges from the Navier--Stokes solution at $a=0$ and exhibits pronounced rod descending behavior for negative values of $a$. In contrast, Model~II behaves continuously across $a=0$, showing no qualitative change in its response and continuing to enhance rod descending progressively. Both models produce the same result at $a=-1$, as can be directly inferred from the governing equations (\Cref{eq:modelA} and \Cref{eq:modelB}). This common limit corresponds to the Oldroyd--A model, or equivalently, the Oldroyd model formulated with the lower-convected derivative.

\begin{figure}[h!]
    \centering
    \begin{subfigure}[b]{0.495\textwidth}
        \centering
        \includegraphics[width=\textwidth]{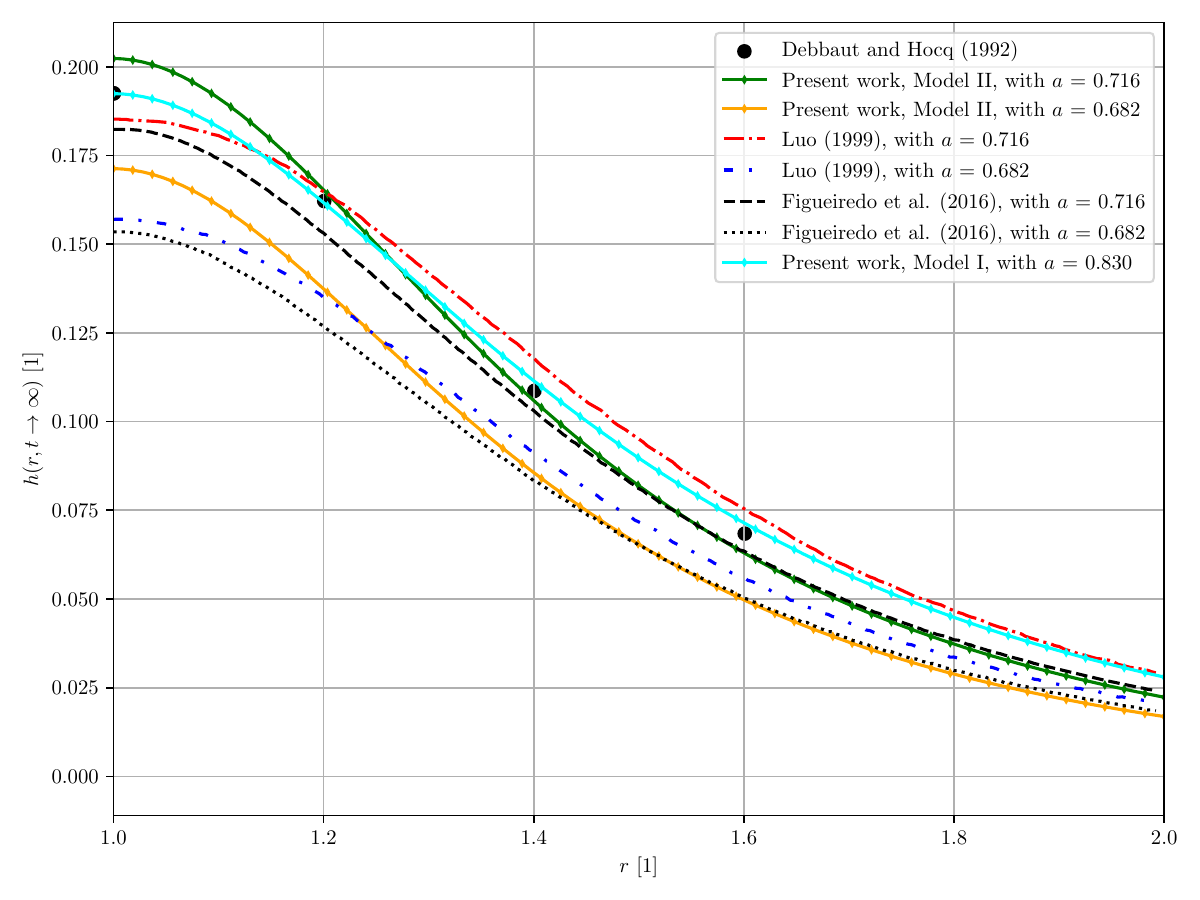}
        \caption{$\omega = 1.7$ rev/s}
        \label{fig:heatmap_A}
    \end{subfigure}
    \hfill
    \begin{subfigure}[b]{0.495\textwidth}
        \centering
        \includegraphics[width=\textwidth]{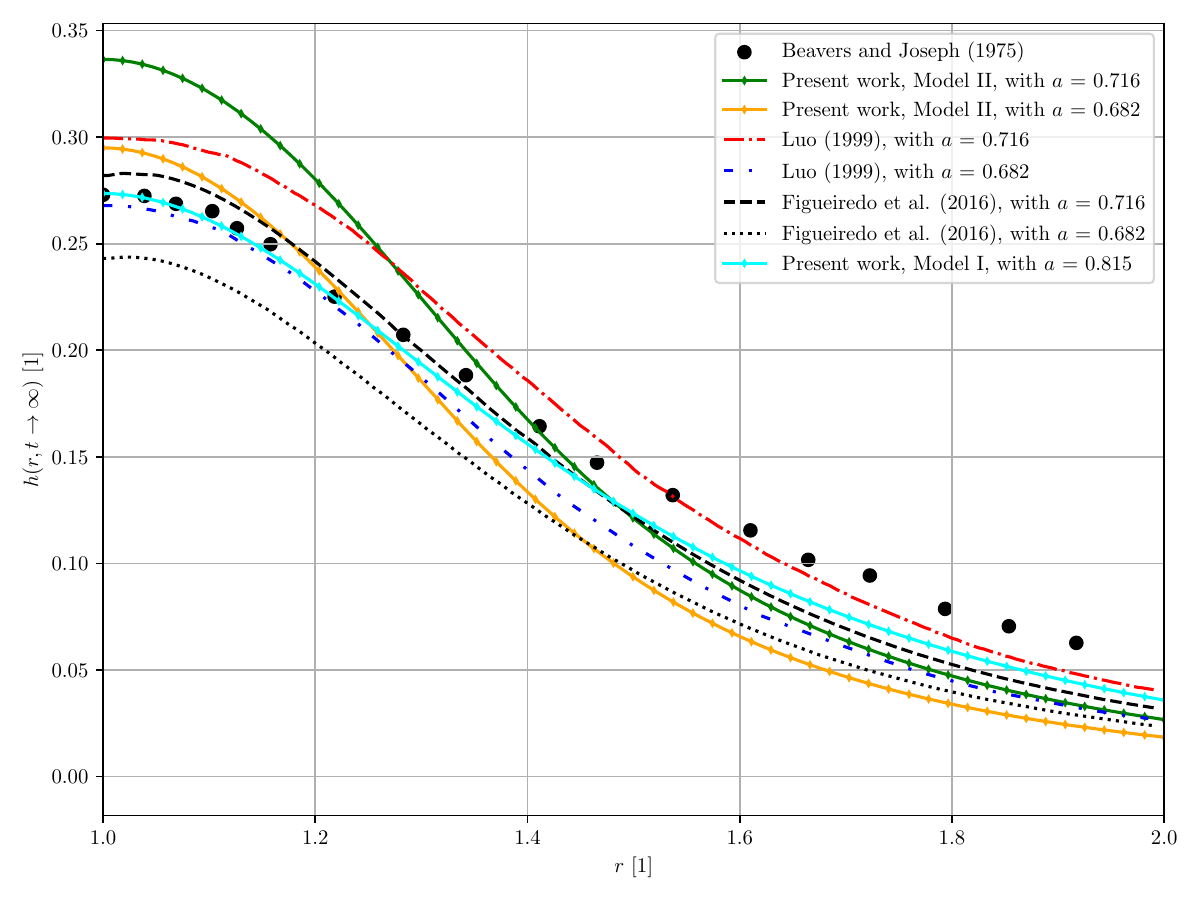}
        \caption{$\omega = 2.1$ rev/s}
        \label{fig:heatmap_B}
    \end{subfigure}

    \vskip\baselineskip %

    \begin{subfigure}[b]{0.495\textwidth}
        \centering
        \includegraphics[width=\textwidth]{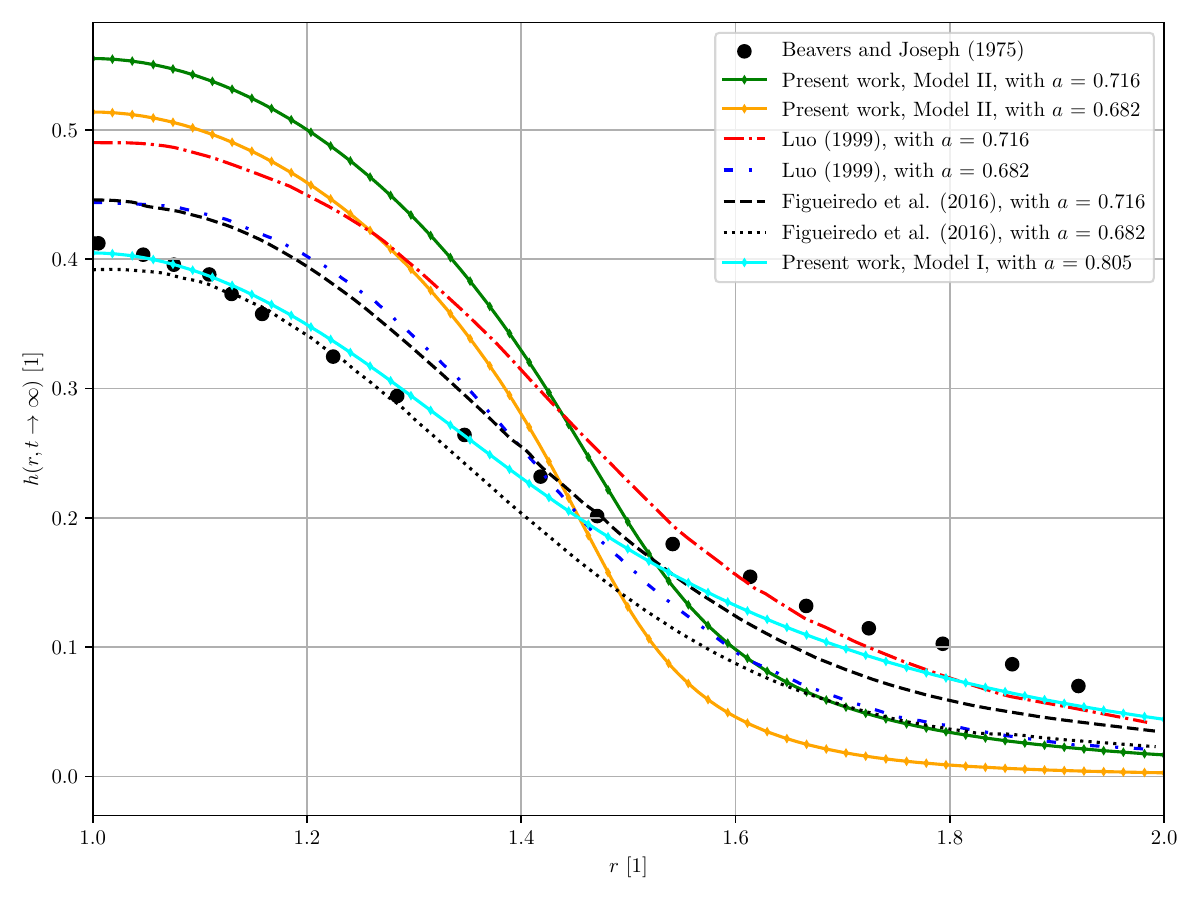}
        \caption{$\omega = 2.6$ rev/s}
        \label{fig:heatmap_C}
    \end{subfigure}
    \hfill
    \begin{subfigure}[b]{0.495\textwidth}
        \centering
        \includegraphics[width=\textwidth]{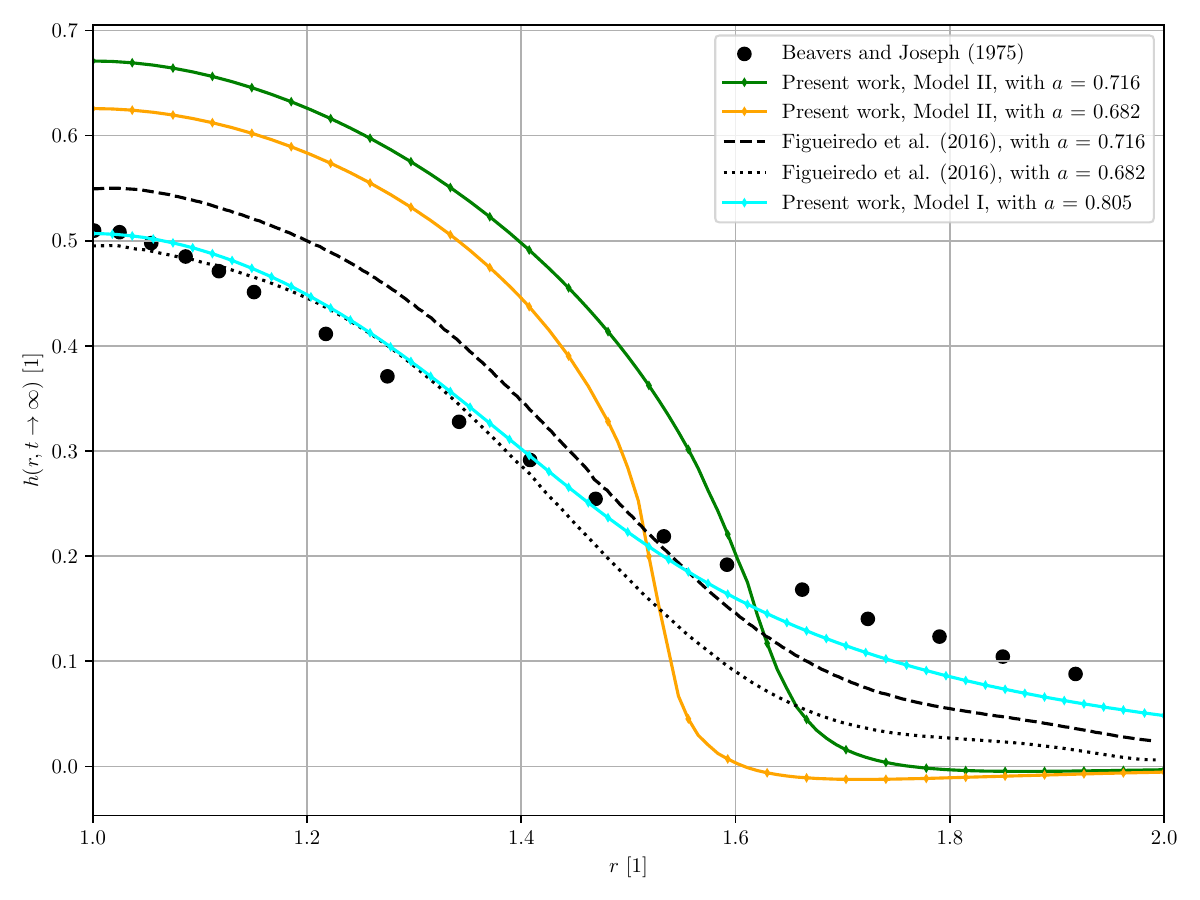}
        \caption{$\omega = 2.9$ rev/s}
        \label{fig:heatmap_D}
    \end{subfigure}

\caption{Comparison of rod climbing predictions for the commonly used Johnson--Segalman model in the engineering community (Model~II) and the thermodynamically consistent Model~I. Experimental data from \cite{Beavers_Joseph_1975,DEBBAUT1992103} are shown for reference and are well captured by Model~I (cyan). Model~II results from our implementation are compared with previous numerical studies \cite{FIGUEIREDO201698, LUO1999393}. All data are extracted by automeris.io \cite{WebPlotDigitizer}. In the left Cauchy--Green tensor $\Bp$ formulation used here, the slip parameter $a$ corresponds to $\xi = 1 - a$, where $\xi$ is the standard slip parameter in the extra-stress $\mathbb{S}$ formulation. In that formulation, the Cauchy stress is written as $\mathbb{T} = -p\mathbb{I} + 2\mu_s \mathbb{D} + \mathbb{S}$, and the Model~II takes the form $\tau \oldroyd{\mathbb{S}} = 2\mu_p \mathbb{D} - \mathbb{S}- \tau \xi (\mathbb{S}\mathbb{D} +\mathbb{D}\mathbb{S})$.}

    \label{fig:JSref_all}
\end{figure}

To further assess the predictive capabilities of the two formulations, we compare their numerical predictions with the available experimental and numerical results in \Cref{fig:JSref_all}. First, we validate our implementation of Model~II against published numerical studies. Similarly to the Oldroyd--B case discussed previously, we observe systematically higher climbing heights than those reported in the literature, likely for the same reasons. Interestingly, our results are closer to those reported in \cite{FIGUEIREDO201698} than to those in \cite{LUO1999393}, despite the fact that our numerical methodology is more closely related to the latter.

More importantly, the thermodynamically consistent Model~I (cyan color in the plot) provides significantly better agreement with the experimental measurements of rod climbing reported in \cite{Beavers_Joseph_1975,DEBBAUT1992103}. In particular, Model~I reproduces both the magnitude and the trend of the climbing height better over a wide range of rotation speeds, including higher angular velocities ($\omega = 2.9~\mathrm{rev/s}$). We note that further quantitative improvement may be achieved by adjusting the relaxation time $\tau$ and the surface tension coefficient $\gamma$, which are among the most difficult material parameters to determine experimentally and are therefore subject to comparatively large uncertainties. We regard this as an opportunity for future work in conjunction with additional experimental results.

In summary, the present comparison indicates that the thermodynamically consistent Johnson--Segalman formulation (Model~I) not only satisfies the second law of thermodynamics by construction, but also yields physically plausible free-surface behavior across the range of the slip parameter $a \in [0,1]$ and provides better agreement with available experimental data. By contrast, the engineering formulation may generate qualitatively anomalous predictions in free-surface flows dominated by normal stresses. These findings support the use of thermodynamically grounded constitutive models in simulations of viscoelastic flows with deformable interfaces.

\subsection{Giesekus model}
To further validate our implementation, in this section we consider the Giesekus model (a special case of the JSG model (\ref{eq:modelA}) with $a=1$, $\alpha \in [0,0.5]$) and reproduce the analytical condition for the presence of fluid climbing using our numerical implementation, achieving good agreement. The paper by Ruangkriengsin et al.~\cite{Ruangkriengsin_Brandão_Wu_Hwang_Boyko_Stone_2025} suggests a simple, sharp criterion distinguishing between descending and climbing regimes of the fluid, highlighting the main physical parameters involved.

They identify the interplay between the Bond number $B$ (surface-tension-to-gravity ratio) and the modified elasticity number $\Lambda = 4 \beta_p (1-2\alpha) \We / \Re$, which consists of the Weissenberg-to-Reynolds number ratio (the original elasticity number) modified by the polymer viscosity ratio $\beta_p$ and the mobility parameter of the Giesekus model $\alpha$. The limit of the Oldroyd-B model seems attainable. The condition they propose is as follows:
\begin{equation}
    \Lambda > \frac{
\int_{1}^{\infty} R^{-1}\, K_0(\sqrt{B}R)\, \mathrm{d}R
}{
\int_{1}^{\infty} R^{-3}\, K_0(\sqrt{B}R)\, \mathrm{d}R
},
\label{eq:condition}
\end{equation}
where $K_0$ is the Bessel $K$ function of zero order.

To reproduce this result, we implement the Giesekus model and choose a range of parameters that satisfy the assumptions of the parameter expansion method used to derive the above condition. Namely, we require $\We \ll 1$, $\Re \ll 1$, and $\mathcal{G} \gg 1$. The last condition corresponds to the aim of ensuring small deformations and is also the limit in which the formula was explicitly obtained. In particular, the condition is derived by order matching under the assumption $\partial_r h(r) \ll 1$, which forbids large slopes of the free surface.

\begin{figure}[h!]
    \centering
    \begin{subfigure}[b]{0.495\textwidth}
        \centering
        \includegraphics[width=\textwidth]{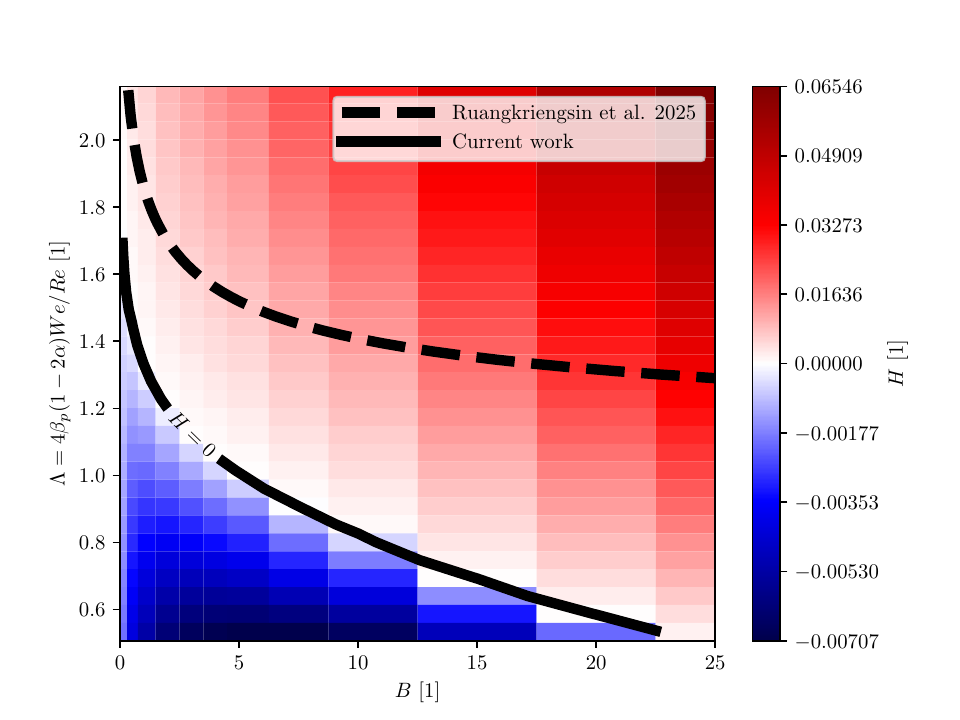}
        \caption{$\We = 0.195$, $\mathcal{G} = 0.069$}
        \label{fig:heatmap_A}
    \end{subfigure}
    \hfill
    \begin{subfigure}[b]{0.495\textwidth}
        \centering
        \includegraphics[width=\textwidth]{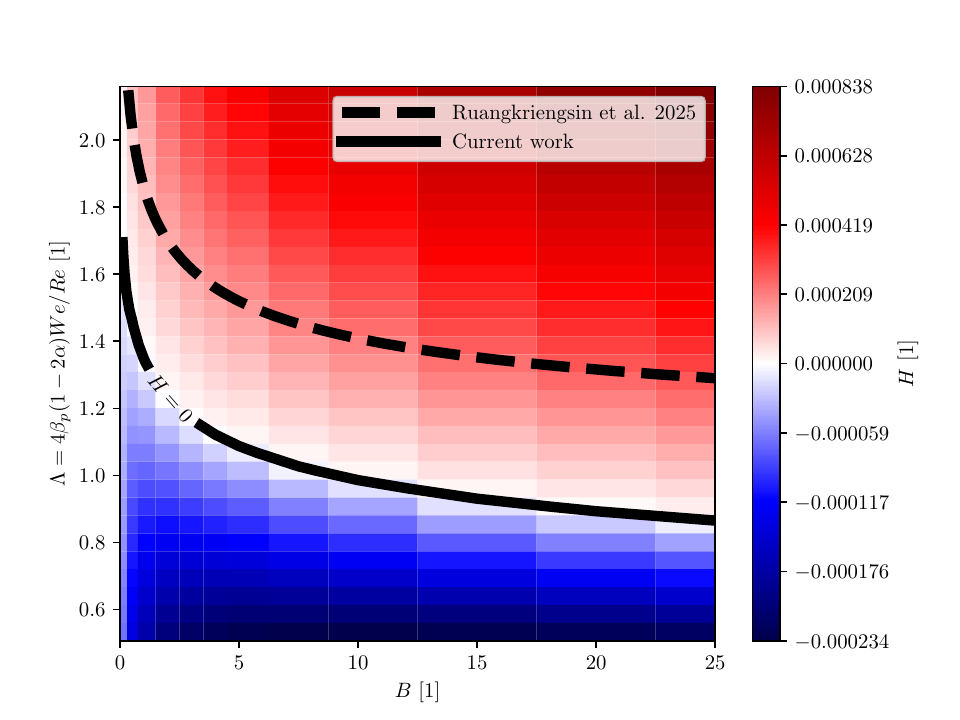}
        \caption{$\We = 0.195$, $\mathcal{G} = 2.076$}
        \label{fig:heatmap_B}
    \end{subfigure}

    \vskip\baselineskip %

    \begin{subfigure}[b]{0.495\textwidth}
        \centering
        \includegraphics[width=\textwidth]{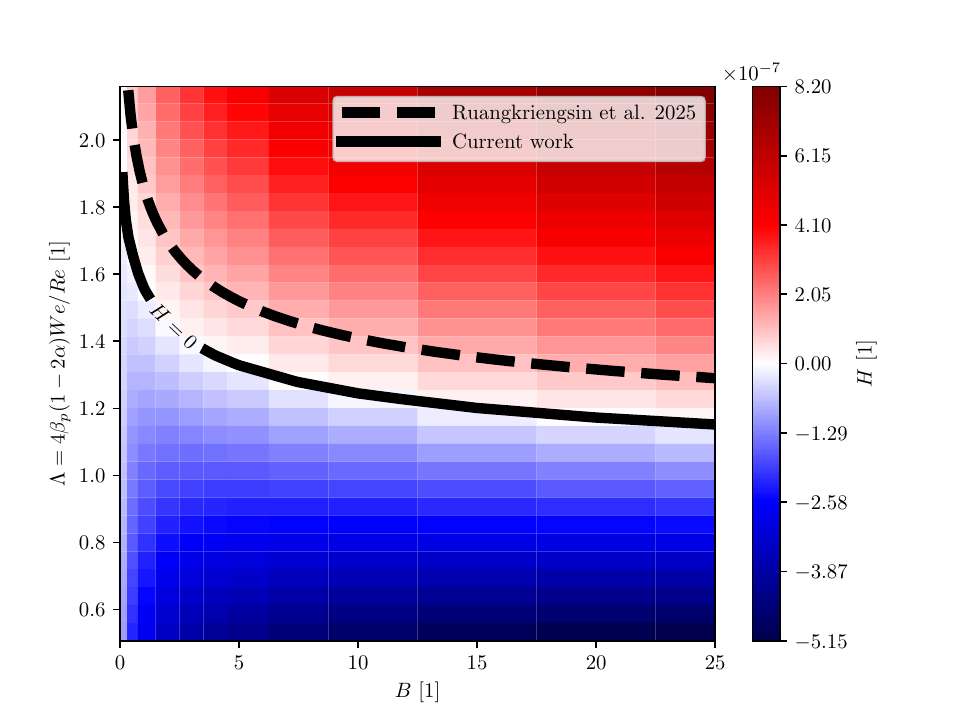}
        \caption{$\We = 0.039$, $\mathcal{G} = 69.2$}
        \label{fig:heatmap_C}
    \end{subfigure}
    \hfill
    \begin{subfigure}[b]{0.495\textwidth}
        \centering
        \includegraphics[width=\textwidth]{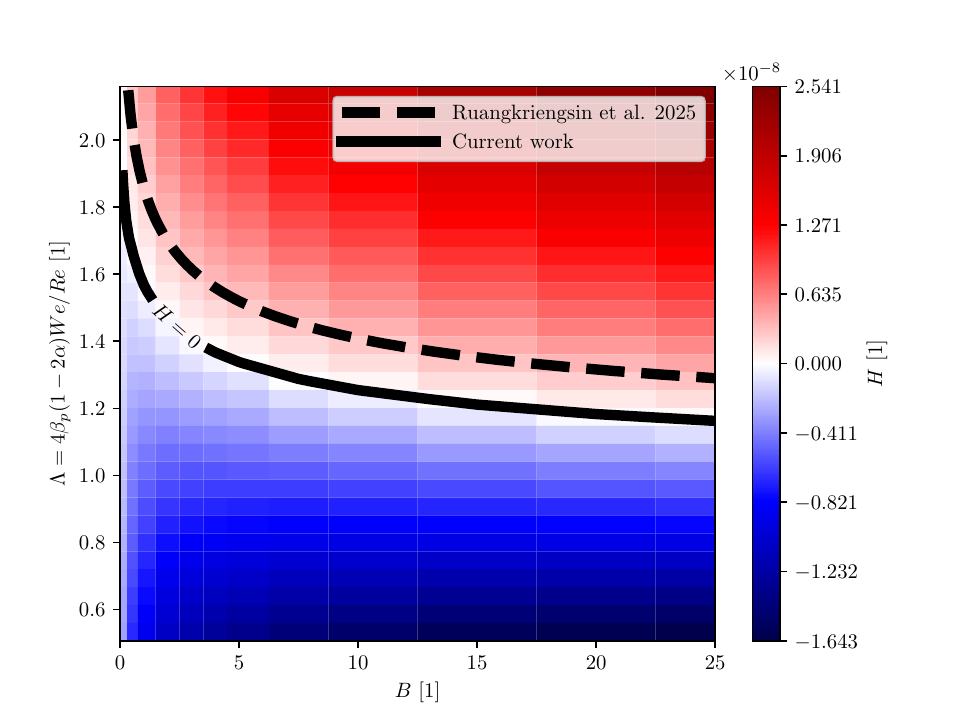}
        \caption{$\We = 9.73 \times 10^{-3} \ll 1$, $\mathcal{G} = 138 \gg 1$}
        \label{fig:heatmap_D}
    \end{subfigure}

    \caption{Rod climbing heat maps for Giesekus model with different parameter sets $(\We,\mathcal{G})$. The dashed line is zero level set of the condition (\ref{eq:condition}). The colorbar is broken at zero, with separate linear scales for positive and negative values.}
    \label{fig:climbing_heatmaps}
\end{figure}

We choose the same setting as in \Cref{sec:configuration}. The only differences are the values of $\omega$, $g$, and $\gamma$. The parameter $\alpha$ is new in this model. By adjusting $\omega$, we satisfy $\We \ll 1$ and $\Re \ll 1$ to arbitrary accuracy. Similarly, adjusting $g$ ensures $\mathcal{G} \gg 1$. For the parametric study, as in the referenced paper \cite{Ruangkriengsin_Brandão_Wu_Hwang_Boyko_Stone_2025}, we scan the intervals $B \in (0.1,25)$ and $\Lambda \in (0.5,2.2)$. This can be done independently of the other parameters by adjusting $\gamma$ and $\alpha$.
In particular, we choose $\omega$ and $g$ such that we obtain the following $\We,\,\mathcal{G}$ pairs, see \Cref{tab:giesekus_cases}. With these values, we obtain the plots shown in \Cref{fig:climbing_heatmaps}.

\begin{table}[h!]
\centering
\caption{Rotation rates and gravity parameters used in the simulations and the corresponding dimensionless numbers for the Giesekus model.}
\label{tab:giesekus_cases}
\begin{tabular}{c c c c c c c}
\toprule
Case & $\omega$ (rev/s) & $g$ (m/s$^2$) & $\We$ & $\mathcal{G}$ & $\Lambda$ & $B$ \\
\midrule
a) & 1.7   & 9.81              & 0.195    & 0.0692   & [0.53, 2.14] & [0.1, 25] \\
b) & 1.7   & $9.81\times30$    & 0.195    & 2.076   & [0.53, 2.14] & [0.1, 25] \\
c) & 0.34  & $9.81\times1000$  & 0.0389   & 69.2     & [0.53, 2.14] & [0.1, 25] \\
d) & 0.085 & $9.81\times2000$  & 0.00973 & 138.4    & [0.53, 2.14] & [0.1, 25] \\
\bottomrule
\end{tabular}
\end{table}

Even though the authors suggest that their results may remain valid beyond the strict asymptotic assumptions, our numerical results show that the analytically predicted climbing condition~\Cref{eq:condition} departs from the numerically observed transition between climbing and descending regimes as these assumptions are relaxed. Within the regime where the assumptions are well satisfied, the analytical and numerical transition curves exhibit the same qualitative shape, differing only by a small systematic quantitative offset. As the parameters move further away from the asymptotic regime, this offset increases, and eventually, the shape of the numerically observed transition curve itself deviates qualitatively from the analytical prediction. On the other hand, the analytical condition~\eqref{eq:condition} consistently overestimates the onset of climbing; consequently, it provides a conservative decision criterion.

Unfortunately, perfect agreement with the analytical solution is not achieved, even when the assumptions of the model appear to be well satisfied. A small but systematic multiplicative offset remains. This discrepancy may arise from deviations from the idealized configuration considered in the analytical study, such as the finite size of the computational domain and the neglect of the free-surface contact condition at the rod. These effects can influence the results, even though the simulation domain is relatively large and the a posteriori evaluated steady-state contact angle is close to $90^\circ$. Since the physical deformation of the domain is very small, even subtle effects can become noticeable.

The discrepancy does not appear to stem from neglecting interactions between the free surface and the ambient phase, since the analytical study similarly accounts only for surface tension at the interface. Ultimately, the remaining offset may also reflect the intrinsic accuracy limits of the asymptotic expansion method itself.

\section{Conclusion} \label{sec:conclusion}

In this work, we introduced and analyzed a class of
Johnson--Segalman--Giesekus (JSG) models combining the
Gordon--Schowalter objective derivative with a Giesekus-type non-linear
relaxation mechanism. Within a thermodynamically consistent framework, we
derived a conformation-tensor-based JSG model from a prescribed Helmholtz
free energy and rate of dissipation, thereby identifying a formulation that
inherently satisfies the second law of thermodynamics. In contrast, the
classical engineering stress-based Johnson--Segalman model, even when
augmented by a Giesekus-type non-linear term, does not generally admit a
non-negative rate of dissipation beyond the upper-convected Oldroyd
limit. This distinction highlights fundamental structural differences
between thermodynamically admissible Johnson--Segalman-type models and their
commonly used engineering counterparts.

We present a numerical code for simulating the rod climbing effect based on
an ALE formulation. The implementation is computationally efficient, with
typical runtimes of at most a few minutes depending on the fluid properties,
and provides accurate predictions of free-surface deformation in regimes
relevant to rotating viscometer experiments. Owing to the limitations of the
ALE description, the framework cannot, in its current form, capture surface
shapes that develop vertical slopes or overhangs without additional
treatment. Nevertheless, it is particularly well suited for the simulation
of standard rheometric configurations, where the free surface remains smooth
and single-valued. Several viscoelastic constitutive models are implemented,
and the framework can be readily extended to additional models.
The code has been released openly for reproducibility and further use \cite{cach_rod_climbing2026}. Finally, our computational framework
provides a natural basis for future comparisons with analytical perturbation
approaches and experimental measurements. Preliminary results further
suggest that the extension to fully three-dimensional computations is also
feasible.

Our numerical results indicate that the thermodynamically consistent
Johnson--Segalman formulation (Model~I) not only respects the second law of
thermodynamics by construction but also leads to phenomenologically
observable free-surface behavior across the full positive range of
the slip parameter. In particular, the improved agreement with available
experimental data supports the use of thermodynamically grounded
constitutive models in simulations of viscoelastic flows with deformable
interfaces.

\acknow
J.C.\ thanks the Charles University Grant Agency for support under Grant No.~131124. J.C.\ was supported by the Ministry of Education, Youth and Sports (MEYS) CR under the project OP JAK CZ.02.01.01/00/22\_008/0004591. P.E.F.~was supported by
the Engineering and Physical Sciences Research Council [grant numbers EP/R029423/1 and EP/W026163/1],
the Science and Technology Facilities Council [grant number UKRI/ST/B000495/1],
the Donatio Universitatis Carolinae Chair ``Mathematical modelling of multicomponent systems'', 
the UKRI Digital Research Infrastructure Programme through the Science and Technology Facilities Council's Computational Science Centre for Research Communities (CoSeC),
and
the Swedish Research Council under grant No.~Z2021-06594 while in residence at Institut Mittag-Leffler in Djursholm, Sweden.
J.M.~and K.T.~have been supported by the project No.~23-05207S financed by the Czech Science Foundation, Czech Republic (GAČR). 
J.C., J.M.~and K.T.~are members of the Charles University Research Centre, Czech Republic program No.~UNCE/24/SCI/005. J.M.~and K.T.~are members of the Nečas Center for Mathematical Modeling.
For the purpose of open access, the authors have applied a CC BY public copyright licence to any author accepted manuscript arising from this submission.
No new data were generated or analysed during this study.

{
\hypersetup{
  colorlinks=false, %
  urlcolor=black,
  linkcolor=black,
  citecolor=black
}
\bibliographystyle{elsarticle-num}

}

\end{document}